\documentclass[12pt,bibnotes,footinbib]{revtex4}
\usepackage{amsmath}
\usepackage{amsfonts}
\usepackage{amssymb}
\usepackage{graphicx}
\usepackage{epstopdf}

\font\msytw=msbm9 scaled\magstep1

\let\a=\alpha \let\b=\beta  \let\g=\gamma  \let\d=\delta \let\e=\varepsilon
  \let\h=\eta   \let\th=\theta  \let\l=\lambda
\let\m=\mu    \let\n=\nu    \let\x=\xi     \let\p=\pi    \let\r=\rho
\let\s=\sigma \let\t=\tau    \let\ph=\varphi\let\c=\chi
   \let\o=\omega
\let\G=\Gamma \let\D=\Delta  \let\L=\Lambda

\def\EE{{\cal E}} \def\VV{{\cal V}}
 \def\WW{{\cal W}}
\def\TT{{\cal T}} \def\BBB{{\cal B}}\def\III{{\cal I}}
\def\RR{{\cal R}}\def\LL{{\cal L}}  
\def\GG{{\cal G}} \def\SS{{\cal S}}

   \def\qq{{\bf q}}
   \def\pp{{\bf p}}
 \def\xx{{\bf x}}  

\def\kk{{\bf k}}

\def\RRR{\hbox{\msytw R}}

 \def\ZZZ{\hbox{\msytw Z}}

\def\\{\hfill\break}
\def\={:=}
\let\io=\infty
\let\0=\noindent
\def\media#1{{\langle#1\rangle}}
\let\dpr=\partial

\def\const{{\rm const}}
\def\tende#1{\,\vtop{\ialign{##\crcr\rightarrowfill\crcr\noalign{\kern-1pt
    \nointerlineskip} \hskip3.pt${\scriptstyle #1}$\hskip3.pt\crcr}}\,}
\def\otto{\,{\kern-1.truept\leftarrow\kern-5.truept\to\kern-1.truept}\,}

\def\to{\rightarrow}

\def\qed{\hfill\raise1pt\hbox{\vrule height5pt width5pt depth0pt}}
\def\Val{{\rm Val}}
\def\ul#1{{\underline#1}}
\def\lis{\overline}
\def\V#1{{\bf#1}}
\def\be{\begin{equation}}
\def\ee{\end{equation}}
\def\bea{\begin{eqnarray}}
\def\eea{\end{eqnarray}}
\def\nn{\nonumber}
\def\pref#1{(\ref{#1})}

\def\lb{\label}

\def\sl{{\displaystyle{\not}}}
\def\Tr{\mathrm{Tr}}

\newtheorem{theorem}{Theorem}[section]

\begin{document}
\title[Anomalous behavior in an effective model of graphene]{Anomalous behavior
in an effective model of graphene with Coulomb interactions}
\author{A. Giuliani}
\affiliation{Universit\`a di Roma Tre, L.go S. L. Murialdo 1,
00146 Roma - Italy}
\author{V. Mastropietro}
\affiliation{Universit\`a di Roma Tor Vergata, V.le della Ricerca
Scientifica, 00133 Roma - Italy}
\author{M. Porta}
\affiliation{Universit\`a di Roma La Sapienza, P.le Aldo Moro 2,
00185 Roma - Italy}
\begin{abstract}
We analyze by exact Renormalization Group (RG) methods the
infrared properties of an effective model of graphene, in which
two-dimensional massless Dirac fermions propagating with a
velocity smaller than the speed of light interact with a
three-dimensional quantum electromagnetic field. The fermionic
correlation functions are written as series in the running
coupling constants, with finite coefficients that admit explicit
bounds at all orders. The implementation of Ward Identities in the
RG scheme implies that the effective charges tend to a line of
fixed points. At small momenta, the quasi-particle weight tends to
zero and the effective Fermi velocity tends to a finite value.
These limits are approached with a power law behavior
characterized by non-universal critical exponents.
\end{abstract}

\maketitle

\renewcommand{\thesection}{\arabic{section}}

\section{Introduction and main result}\label{sec1}
\setcounter{equation}{0}
\renewcommand{\theequation}{\ref{sec1}.\arabic{equation}}

The charge carriers in graphene at half filling are effectively
described by massless Dirac fermions constrained to move on a
two-dimensional (2D) manifold embedded in three-dimensional (3D)
space \cite{C}, with a Fermi velocity $v$ that is approximately
300 times smaller than the speed of light. As a consequence,
already without taking into account the interactions, the system
displays highly unusual features as compared to standard 2D electron gases,
such as an anomalous integer
quantum Hall effect and the insensitivity to disordered-induced
localization; most of these effects have already been
experimentally observed \cite{N2,N1}. The study of many-body
interactions among the charge carriers in graphene is of course
very important, particularly in view of recent experiments that
suggest their relevant role in several physical properties of
graphene \cite{B1,JHTWSHKS07,LLA09,ZSFL08}.

The effect of a weak {\it short} range interaction in graphene is
quite well understood: it turns out that the behavior of the
ground state is qualitatively similar to the free one, except that
the Fermi velocity and the wave function renormalization are
renormalized by a finite amount. This was expected on the basis of
a power counting analysis \cite{V2,H,HJR}; recently, it has been
rigorously proven in \cite{GM}, where the {\it convergence} of the perturbative
series was established, by using the methods of
constructive Quantum Field Theory (QFT) and by taking into full account
the lattice effects (i.e., by considering the Hubbard model
on the honeycomb lattice).

The situation in the presence of {\it long} range interactions is
much more subtle and still not completely understood. Their effect
in graphene is usually studied in terms of a model of Dirac
fermions interacting via a {\it static} Coulomb potential;
retardation effects of the electromagnetic (e.m.) field are
neglected because the free Fermi velocity $v$ is 300 times smaller
than the speed of light $c$. At weak coupling, a {\it logarithmic
divergence} of the effective Fermi velocity $v(\kk)$ at the Fermi
points $\pp_F^\pm$ and a finite quasi-particles weight have been
predicted, on the basis of one-loop \cite{V3} and two-loops
\cite{Mis} computations. An unbounded growth of the effective
Fermi velocity was also confirmed by an analysis based on a
large-$N$ expansion \cite{KUC,S}, which predicted a {\it power
law} divergence of $v(\kk)$ at $\pp_F^\pm$. It is not clear how to
reconcile the logarithmic divergence expected from two-loops
perturbative computations with the power law behavior found by
large-$N$ expansions; moreover, the description in terms of Dirac
fermions introduces spurious ultraviolet divergences that can
produce ambiguities in the physical predictions \cite{HJR,Mis}.
These difficulties may be related to a basic inadequacy of the
effective model of Dirac fermions with static Coulomb
interactions: the fact that the effective Fermi velocity diverges
at the Fermi points (as predicted by all the analyses of the
model) signals that its physical validity {\it breaks down} at the
infrared scale where $v(\kk)$ becomes comparable with the speed of
light; at lower scales, retardation effects must be taken into
account, as first proposed in \cite{V1}, where a model of massless
Dirac fermions propagating with speed $v\ll c$ and interacting
with an e.m. field was considered. In \cite{V1} it was found that,
at small momenta, the wave function renormalization diverges as a
power law; this implies that the ground state correlations have an
anomalous decay at large distances. Moreover, it was found that
the interacting Fermi velocity increases up to the speed of light,
again with an anomalous power law behavior. Despite its interest,
the model proposed in \cite{V1} has not been considered further.
The results in \cite{V1} were found on the basis of one-loop
computations and in the presence of an ultraviolet dimensional
regularization scheme. It is interesting to investigate whether
the predictions of \cite{V1} remain valid even if higher orders
corrections are taken into account and in the presence of
different regularization schemes closer to the lattice cut-off
that is truly present in actual graphene.

The model we consider describes massless Dirac fermions in $2+1$
dimensions propagating with velocity $v<c$, and interacting with a
$3+1$ dimensional photon field in the Feynman gauge. We will not
be concerned with the instantaneous case ($c\to\io$); therefore,
from now on, for notational simplicity, we shall fix units such
that $\hbar=c=1$. The model is very similar to the one in
\cite{V1}, the main difference being the choice of the ultraviolet
cut-off: rather than considering dimensional regularization, in
order to mimic the presence of an underlying lattice, we
explicitly introduce a (fixed) ultraviolet momentum cutoff both in
the electronic and photonic propagators. The correlations can be
computed in terms of derivatives of the following euclidean
functional integral:
\be e^{\WW(J,\phi)} = \int P(d\psi)P(dA)e^{V(A,\psi) +B(J,\phi)}
\lb{1.1} \ee
with, setting $\xx = (x_0,\vec x)$ and $\vec x=(x_1,x_2)$
(repeated indexes are summed; greek and latin labels run
respectively from $0$ to $2$, $1$ to $2$),
\bea &&V(A,\psi):= \int_{\L} d\xx\, \big[e\,j_{\m,\xx}A_{\m,\xx} -
\n_{\m}A_{\m,\xx}A_{\m,\xx}\big]\lb{1.2}\\&& B(J,\phi):= \int_{\L}
d\xx\, \big[j_{\m,\xx}J_{\m,\xx} + \phi_{\xx}\lis\psi_{\xx} +
\lis\phi_{\xx}\psi_{\xx}\big]\;,\qquad\nn \eea
where $\L$ is a three dimensional box of volume $|\L| = L^{3}$
with periodic boundary conditions (playing the role of an infrared
cutoff, to be eventually removed), the couplings $e$, $\n_{\m}$
are real and $\n_{1} = \n_{2}$; the couplings $\n_\m$ are {\it
counterterms} to be fixed so that the photon mass is vanishing in
the deep infrared. Moreover, $\lis\psi_{\xx}$, $\psi_{\xx}$ are
$4$-components {\em Grassmann spinors}, and the $\m$-th component
$j_{\m,\xx}$ of the current is defined as:
\be j_{0,\xx} = i\lis\psi_{\xx}\g_{0}\psi_{\xx}\;,\qquad \vec
j_{\xx} = i v\,\lis\psi_{\xx}\vec \g\psi_{\xx}\;,\lb{1.3} \ee
where $\g_\m$ are euclidean gamma matrices, satisfying the
anticommutation relations $\{\g_\m,\g_\n\}=-2\d_{\m,\n}$. The
symbol $P(d\psi)$ denotes a Grassmann integration with propagator
\be g^{(\le 0)}(\xx) := \int \frac{d\kk}{(2\pi)^3}\, e^{i\kk\xx}
\frac{i k_0 \g_0+i v\vec{k}\cdot\vec{\g}}{ k_0^2+v^2 |\vec
k|^2}\chi_{0}(\kk)\;. \lb{1.4}\ee
where $(2\pi)^{-3}\int d\kk$ is a shorthand for
$|\L|^{-1}\sum_{\kk=2\p{\bf n}/L}$ with ${\bf n}\in\ZZZ^3$, and
$\chi_0(\kk)=\chi(|\kk|)$ plays the role of a prefixed ultraviolet
cutoff (here $\chi(t)$ is a non increasing $C^{\infty}$ function
from $\RRR^{+}$ to $[0,1]$ such that $\chi(t) = 1$ if $t\leq 1$
and $\chi (t) = 0$ if $t\geq M>1$). Finally, $A_{\m,\xx}$ are
gaussian variables and $P(dA)$ is a gaussian integration with
propagator
\be w^{(\le 0)}(\xx) := \int \frac{d\pp}{(2\pi)^3}\, e^{i\pp\xx}
\frac{\chi_{0}(\pp)}{2|\pp|} = \int \frac{d\pp\, dp_3}{(2\pi)^4}\,
e^{i\pp\xx} \frac{\chi_{0}(\pp)}{\pp^2 + p_3^{2}}\;.\label{1.5}
\ee
We perform an analysis based on the methods of {\it constructive
Renormalization Group} (RG) for non relativistic fermions,
introduced in \cite{BG, FT} (see \cite{BG1, M, R, Sal} for updated
introductions), which have already been proved effective in the
study of several low-dimensional critical systems, such as
one-dimensional (1D) interacting fermions \cite{BG,BGPS,BM}, 
2D critical Ising and vertex models \cite{BFMprl,GMprl}, the 2D Hubbard
model on the square lattice at positive temperatures \cite{DR},
interacting fermions with asymmetric Fermi surfaces \cite{FKT} and
the 2D Hubbard model on the honeycomb lattice \cite{GM}, just to
mention a few. Compared to other RG approaches, such as those in
\cite{P, Sh}, the advantage of the constructive methods we adopt
is that they allow us to get a rigorous and complete treatment of
the effects of the cut-offs and a full control on the perturbative
expansion via explicit bounds at all orders; quite remarkably, in
certain cases, such as the ones treated in \cite{BM, DR, GM},
these methods even provide a way to prove the {\it convergence} of
the resummed perturbation theory.

By using these methods, we construct a renormalized expansion,
allowing us to express the Schwinger functions, from which the
physical observables can be computed, as series in the effective
couplings (the effective charges and the effective photon masses, also called 
in the following the {\it running coupling constants}), with {\it finite}
coefficients at all orders, admitting explicit $N!$ bounds (see
Theorem \ref{thm1} in Section \ref{sec3} below). If the effective
couplings remain small in the infrared, informations obtained from
our expansion by lowest order truncations are reliable at weak
coupling. The importance of having an expansion with finite
coefficients should not be underestimated; the naive perturbative
expansion in the fine structure constant is plagued by {\it
logarithmic infrared divergences} and higher orders are more and
more divergent.

Of course the renormalized expansion is useful only as long as the
running coupling constants are small. In fact, we do prove that
they remain small for all infrared scales, by implementing Ward
Identities (WIs) in the RG flow, using a technique developed in
\cite{BM} for the rigorous analysis of Luttinger liquids in
situations where bosonization cannot be applied (e.g., in the
presence of an underlying lattice and/or of non-linear bands). The
WIs that we use are based on an approximate local gauge
invariance, the exact gauge symmetry being broken by the
ultraviolet cut-off; its presence produces corrections to the
``naive'' (formal) WIs, which can be resummed and, again,
explicitly bounded at all orders. The resulting modified WIs imply
that the effective charges tend to a line of fixed points, exactly
as in 1D Luttinger liquids. We note that this is one of the very
few examples in which Luttinger liquid behavior is found in
dimensions higher than $1$.

Let us denote by $\media{\ldots}=\lim_{|\L|\to\io}\media{\ldots}_\L$ the
expectation value with respect to the interaction (\ref{1.2}) in
the infinite volume limit; our main result can be informally
stated as follows (more rigorous statements will be found below).

\vskip.5cm {\bf Main result.} {\it There exists a choice of
$\n_\m$ such that, for $\kk$ small,
\be \media{\psi_\kk\lis\psi_\kk} = \frac{1}{Z(\kk)}\frac{ik_0
\g_0+i v(\kk)\vec{k}\cdot\vec{\g}}{ k_0^2+v(\kk)^2 |\vec k|^2}(1+B(\kk))\;,
\lb{1.6}\ee
where:
\be Z(\kk) \sim |\kk|^{-\h}\;, \qquad\quad v_{eff} - v(\kk) \sim(v_{eff}-v)
|\kk|^{\tilde \h}\;,\lb{1.7} \ee
$B(\kk),\n_\m,\h,\tilde \h,v_{eff}$ are expressed by series in the
effective couplings with finite coefficients that admit $N!$-bounds
at all orders. Moreover: (i) the first non-trivial contribution to $B(\kk)$ 
is of second order in $e$; (ii)
the first non-trivial contribution to $\n_\m$ is of second order in $e$
and positive; (iii) the first non-trivial contributions to $\h,\tilde\h,v_{eff}$ 
are, respectively:
\be \h^{(2)}=\frac{e^2}{12\p^2}\;,\qquad \tilde\h^{(2)}=\frac{2e^2}{5\p^2}\;,\qquad
v_{eff}^{(2)}=1-F(v)\frac{e^2}{6\p^2}\;,\label{second}\ee
with
\be F(v)=\frac{5}{8}\Big[\big(\frac1{2v^2}-2\big)
\frac{\x_0-\arctan\x_0}{\x_0^3}
+\frac1{2v^2}\frac{\arctan\x_0}{\x_0}\Big]\;,
\qquad \x_0:=\frac{\sqrt{1-v^2}}{v}\;.\label{1.88}\ee
}
\vskip.4cm Note that the theory is not Lorentz
invariant, because $v\not=c$; moreover, gauge symmetry is broken
by the presence of the ultraviolet momentum cut-off. These two
facts produce unusual features as compared to standard QFT models.
In particular, the momentum cut-off produces correction terms in
the WIs, which can be rigorously bounded at all orders; despite
these corrections, one can still use the WIs to prove that the
beta function for the effective charges is asymptotically
vanishing at all orders, so that the model admits a line of
(non-trivial) fixed points.

The lack of gauge invariance due to the ultraviolet momentum
cut-off makes it necessary (as in \cite{BDM}) to introduce
positive counterterms to keep the photon mass equal to zero.
Similarly, it implies that the effective couplings with the
temporal and spatial components of the gauge field are different
and that the effective Fermi velocity $v_{eff}$ is not equal to the
speed of light. However, it is possible to introduce in the bare
interaction two different charges, $e_0$ and $e_1$, describing the
couplings of the photon field with the temporal and spatial
components of the current, which can be tuned so that the dressed
charges are equal and $v_{eff}=1$.

A more realistic model for single layer graphene could be obtained
by considering tight binding electrons hopping on the honeycomb
lattice, whose lattice currents are coupled to a 3D photon field.
A Renormalization Group analysis similar to the one in the present
paper could be repeated for the lattice model, by extending the
formalism in \cite{GM}. If the lattice model is chosen in such a way 
that lattice gauge invariance is preserved, we expect that its photon mass 
counterterms are exactly zero and that its effective Fermi velocity is 
equal to the speed of light. In any case (i.e., both in the presence or 
in the absence of lattice gauge invariance), we expect the lattice model 
to have the same infrared asymptotic behavior of the continuum model 
considered here, provided that the bare parameters $e_\m,\n_\m$ 
of the continuum model are properly tuned. 

Finally, let us comment about the possibility of providing 
a full {\it non perturbative} construction of the ground state
of the present model or, possibly, of a more realistic model of 
tight binding electrons hopping on the honeycomb lattice and interacting with e.m. forces.
In this paper, we express the physical obervables in
terms of series in the running coupling constants (with bounded
coefficients at all orders) and we show that the running coupling
constants remain close to their initial value, thanks to Ward
Identities and cancellations in the beta function. Thus, the
usual problem that so far prevented the non-perturbative
construction of the ground state of systems of interacting fermions 
in $d>1$ with convex symmetric Fermi surface (namely, the presence of a beta 
function driving the infrared flow
for the effective couplings out of the weak coupling regime)
is absent in the present case. Therefore, a full non-perturbative 
construction of the ground state of the present model appears to be 
feasible, by using determinant bounds for the fermionic sector and
cluster expansion techniques for the bosonic sector. Of course, the 
construction is expected to be much more difficult than the one 
in \cite{BM} or \cite{GM}, due to the simultaneous presence of bosons 
and fermions; if one succeeded in providing it, it would represent 
the first rigorous example of anomalous Luttinger-liquid behavior 
in more than one dimension.

The paper is organized as follows: in Section \ref{sec3} we
describe how to evaluate the functional integrals defining the
partition function and the correlations of our model in terms of
an exact RG scheme (details are discussed in Appendix \ref{appsim} and \ref{app3});
in Section \ref{sec3a} we describe the infrared flow of the
effective couplings and prove the emergence of an effective Fermi
velocity different from the speed of light (the explicit lowest
order computations of the beta function are presented in Appendix
\ref{app2}); in Section \ref{sec5} we derive the Ward Identity
allowing us to control the flow of the effective charges (details
are discussed in Appendix \ref{app4}) and proving that the beta
function for the charges is asymptotically vanishing; finally, in
Section \ref{sec6} we draw the conclusions.

\section{Renormalization Group analysis}\label{sec3}
\setcounter{equation}{0}
\renewcommand{\theequation}{\ref{sec3}.\arabic{equation}}
\subsection{The effective potential} In this section we show how to evaluate
the functional integral (\ref{1.1}); the integration will be
performed in an iterative way, starting from the momenta ``close''
to the ultraviolet cutoff moving towards smaller momentum scales.
At the $n$-th step of the iteration the functional integral
(\ref{1.1}) is rewritten as an integral involving only the momenta
smaller than a certain value, proportional to $M^{-n}$, where $M>1$
is the same constant (to be chosen sufficiently close to $1$) 
appearing in the definition of the cut-off function (see lines after (\ref{1.4})), 
and both the propagators and the interaction will be replaced by ``effective''
ones; they differ from their ``bare'' counterparts because the
physical parameters appearing in their definitions (the Fermi
velocity $v$, the charge $e$, and the ``photon mass'' $\n_\m$) are
{\em renormalized} by the integration of the momenta on higher
scales. In the following, it will be convenient to introduce the {\it scale label} $h\leq 0$ as 
$h:=-n$.

Setting $\chi_{h}(\kk):=
\chi(M^{-h}|\kk|)$, we start from the following identity:
\be \chi_0(\kk)=\sum_{h=-\io}^0 f_h(\kk)\;,\quad\quad
f_h(\kk):=\chi_h(\kk)-\chi_{h-1}(\kk)\;;\lb{3.1} \ee
let $\psi = \sum_{h=-\io}^{0}\psi^{(h)}$ and $A = \sum_{h=-\io}^{0}A^{(h)}$,
where $\{\psi^{(h)}\}_{h\le 0}$, $\{A^{(h)}\}_{h\le 0}$ are independent
free fields with the same support of the functions $f_{h}$ introduced above.

We evaluate the functional integral (\ref{1.1}) by integrating the fields in
an iterative way starting from $\psi^{(0)}$, $A^{(0)}$; for simplicity, we
start by treating the case $J=\phi=0$. We define
$\VV^{(0)}(A,\psi):=V(A,\psi)$ and we want to inductively prove that after the
integration of $\psi^{(0)},A^{(0)},\ldots,\psi^{(h+1)},A^{(h+1)}$
we can rewrite:
\be e^{\WW(0,0)} = e^{|\L| E_{h}}\int
P(d\psi^{(\le h)})P(dA^{(\le h)}) e^{\VV^{(h)}(A^{(\le
h)},\sqrt{Z_{h}}\psi^{(\le h)})}\;,\lb{3.3}\ee
where $P(d\psi^{(\le h)})$ and $P(d A^{(\le h)})$ have propagators
\be g^{(\le h)}(\kk) = \frac{\chi_{h}(\kk)} {\tilde
Z_h(\kk)}\frac{ i \g_0 k_0+ i\tilde v_h(\kk) \vec{k}\cdot
\vec{\g}}{k_{0}^{2}+ \tilde v_h(\kk)^2|\vec{k}|^{2}}\;,\qquad
w^{(\le h)}(\pp) = \frac{\chi_{h}(\pp)}{2|\pp|}\;,\label{3.4}
\ee
$\VV^{(h)}$ has the form
\bea &&\VV^{(h)}(A,\psi) = \sum_{\substack{n,m \geq 0\\ n+m\geq
1}}\sum_{\ul{\r},\ul{\m}}\int \Big[\prod_{i=1}^{2n}\frac{d\kk_i}{(2\pi)^3}\Big]
\Big[\prod_{j=1}^{m}\frac{d\pp_j}{(2\pi)^3}\Big]
\,\prod_{i=1}^{n}\lis\psi_{\kk_{2i-1},
\r_{2i-1}}\psi_{\kk_{2i},\r_{2i},}\cdot\nn\\&&\hskip1cm\cdot
\prod_{i=1}^{m}A_{\m_i, \pp_i}
W^{(h)}_{m,n,\ul\r,\ul\m}(\{\kk_i\},\{\pp_j\})\d\left(\sum_{j=1}^{m}\pp_j
+ \sum_{i=1}^{2n}(-1)^{i}\kk_i\right)\;,\label{3.5} \eea
and $E_{h}$, $\tilde Z_{h}(\kk)$, $\tilde v_{h}(\kk)$ and the
kernels $W^{(h)}_{m,n,\ul\r,\ul\m}$ will be defined
recursively.

In order to inductively prove (\ref{3.3}), we split $\VV^{(h)}$ as
$\LL \VV^{(h)} + \RR \VV^{(h)}$, where $\RR =
1-\LL$ and $\LL$, the {\em localization operator}, is a linear
operator on functions of the form (\ref{3.5}), defined by its action on the
kernels $
W^{(h)}_{m,n,\ul\r,\ul\m}$ in the following way:
\bea &&\mathcal{L}W^{(h)}_{0,1,\ul\rho}(\kk) :=
W^{(h)}_{0,1,\ul\rho}({\bf 0}) +
\kk\partial_{\kk}W^{(h)}_{0,1,\ul\rho}({\bf 0})\;,\label{3.6}\\
&&\mathcal{L}W_{1,1,\ul\rho,\mu}^{(h)}(\pp,\kk)
:= W_{1,1,\ul\rho,\mu}^{(h)}({\bf 0},{\bf 0})\;,\nn\\
&&\mathcal{L}W_{2,0,\ul\mu}^{(h)}(\pp) :=
W_{2,0,\ul\mu}^{(h)}({\bf 0}) + \pp\partial_{\pp}
W_{2,0,\ul\mu}^{(h)}({\bf 0})\;,\quad \mathcal{L}
W_{3,0,\ul\mu}^{(h)}(\pp_1,\pp_2) := W_{3,0,\ul\mu}^{(h)}({\bf
0},\V0)\;,\nn\eea
and $\mathcal{L}W_{P}^{(h)} := 0$ otherwise. As it will be clear from the 
dimensional analysis performed in Section \ref{bound} below, these are 
the only terms that need renormalization; in particular, 
$\LL W^{(h)}_{0,1,\ul\rho}(\kk)$ will contribute to the wave 
function renormalization and to the effective Fermi velocity, 
$\LL W^{(h)}_{2,0,\ul\mu}(\pp)$ to the effective photon mass, and 
$\mathcal{L}W_{1,1,\ul\rho,\mu}^{(h)}(\pp,\kk)$ to the effective charge.

As a consequence of the
 symmetries of our model, see Appendix \ref{appsim}, it turns out that
\bea &&W^{(h)}_{1,0,\m}(\V0) = 0\;,\quad W_{3,0,\ul
\mu}^{(h)}({\bf 0},{\bf 0}) =0\;,\quad
W^{(h)}_{0,1,\ul\rho}({\bf 0})= 0\;,\nn\\
&& \hat W^{(h)}_{2,0,\ul \mu}(\V0) =
- \delta_{\mu_{1},\mu_{2}}M^{h}\n_{\m_1,h}\;,\quad \partial_{\pp}\hat
W_{2,0,\ul\mu}^{(h)}({\bf 0})=0\;\qquad\label{3.7} \eea
and, moreover, that
\bea
&&\lis\psi_{\kk}\,\kk\partial_{\kk}
W^{(h)}_{0,1}(\V0)\psi_{\kk} = -iz_{\m,h}k_\m \lis{\psi}_{\kk}\g_{\m}\psi_{\kk}
\label{3.8}\\
&& \lis\psi_{\kk+\pp}W^{(h)}_{1,1,\m}(\V0,\V0) \psi_{\kk}A_{\m,\pp}= i\l_{\m,h}
\lis{\psi}_{\kk+\pp}\g_{\m}\psi_{\kk}A_{\m,\pp}\;,\nn \eea
with $z_{\m,h}$, $\l_{\m,h}$ real, and $z_{1,h} = z_{2,h}$, $\l_{1,h} =
\l_{2,h}$. We can {\em renormalize} $P(d\psi^{(\le h)})$ by adding to the
exponent of its gaussian weight
the local part of the quadratic terms in the fermionic fields; we get that
\bea &&\int P(d\psi^{(\le h)})P(d A^{(\le h)})
e^{\VV^{(h)}(A,\sqrt{Z_{h}}\psi)} =\label{3.9}\\&&\hskip3cm= e^{|\L| t_h} \int
\widetilde P(d\psi^{(\le h)}) P(d A^{(\le h)})
e^{\widetilde \VV ^{(h)}(A,\sqrt{Z_{h}}\psi)}\;,\nn \eea
where $t_h$ takes into account the different normalization of the
two functional integrals, $\widetilde \VV^{(h)}$ is given by
\bea
\widetilde\VV^{(h)}(A,\psi) &=& \VV^{(h)}(A,\psi) + \int \frac{d\kk}{(2\pi)^3}
\,
iz_{\m,h}k_\m \lis{\psi}_{\kk}\g_{\m}\psi_{\kk}\nn\\&=:&
\VV^{(h)}(A,\psi) - \LL_\psi\VV^{(h)}(A,\psi)\;,\label{3.10}\eea
and $\widetilde P(d\psi^{(\le h)})$ has propagator equal to
\be \tilde g^{(\le h)}(\kk) = \frac{\chi_{h}(\kk)} {\tilde
Z_{h-1}(\kk)}\frac{ i \g_0 k_0+ i\tilde v_{h-1}(\kk) \vec{k}\cdot
\vec{\g}}{k_{0}^{2}+ \tilde
v_{h-1}(\kk)^2|\vec{k}|^{2}}\;,\label{3.11} \ee
with
\bea &&\tilde Z_{h-1}(\kk) = \tilde Z_{h}(\kk) +
Z_{h}z_{0,h}\chi_{h}(\kk)\;,\label{3.12}\\&&\tilde Z_{h-1}(\kk)\tilde
v_{h-1}(\kk) = \tilde Z_{h}(\kk)\tilde v_{h}(\kk) +
Z_{h}z_{1,h}\chi_{h}(\kk)\;.\nn \eea
After this, defining $Z_{h-1} := \tilde Z_{h-1}(\V0)$,
we {\em rescale} the fermionic field so that
\be \widetilde \VV^{(h)}(A,\sqrt{Z_h}\psi) = \hat
\VV^{(h)}(A,\sqrt{Z_{h-1}}\psi)\;;\label{3.13} \ee
therefore, setting
\be
v_{h-1}:= \tilde v_{h-1}(\V0)\;,\quad e_{0,h}:=
\frac{Z_{h}}{Z_{h-1}}\l_{0,h}\;,\quad e_{1,h}v_{h-1}=e_{2,h}v_{h-1}:=
\frac{Z_{h}}{Z_{h-1}}\l_{1,h}\;,\label{3.13b}
\ee
we have that:
\be \LL \hat \VV^{(h)}(A^{(\le h)},\sqrt{Z_{h-1}}\,\psi^{(\le h)})=
 \int_\L d\xx\,\Big(Z_{h-1} e_{\m,h} j_{\m,\xx}^{(\le h)}A_{\m,\xx}^{(\le h)}
- M^h\n_{\m,h}A_{\m,\xx}^{(\le h)}A_{\m,\xx}^{(\le h)}
\Big) \label{3.14}\;, \ee
where
\be j_{0,\xx}^{(\le h)}:=i\lis\psi^{(\le h)}_{\xx}\g_0\psi^{(\le h)}_\xx\;,
\qquad \vec j_{\xx}^{(\le h)}:=iv_{h-1}\,\lis\psi^{(\le h)}_{\xx}\vec \g
\psi^{(\le h)}_\xx\;.\label{3.15b}\ee
After this rescaling, we can rewrite (\ref{3.9}) as
\bea && \int P(d\psi^{(\le h)})P(d A^{(\le h)})
e^{\VV^{(h)}(A,\sqrt{Z_{h}}\psi)} = e^{|\L| t_h}
\int P(d\psi^{(\le h-1)}) P(d A^{(\le h-1)})\cdot
\nn\\
&&\hskip1cm\cdot
\int P(d\psi^{(h)}) P(d A^{(h)})
e^{\hat \VV ^{(h)}(A^{(\le h-1)}+A^{(h)}
,\sqrt{Z_{h-1}}(\psi^{(\le h-1)}+\psi^{(h)}))}\;,\label{3.14a}
\eea
where $\psi^{(\le h-1)},A^{(\le h-1)}$ have propagators given by (\ref{3.4})
(with $h$ replaced by $h-1$) and $\psi^{(h)},A^{(h)}$
have propagators given by
\bea &&\frac{g^{(h)}(\kk)}{Z_{h-1}} = \frac{\tilde f_h(\kk)}{Z_{h-1}}\frac{ i
\g_0 k_0+ i\tilde v_{h-1}(\kk) \vec{k}\cdot \vec{\g}}{k_{0}^{2}+
\tilde v_{h-1}(\kk)^2|\vec{k}|^{2}}\,,\qquad w^{(h)}(\pp) =
\frac{f_h(\pp)}{2|\pp|}\;,\nn\\
&&f_h(\kk) = \chi_{h}(\kk) - \chi_{h-1}(\kk)\;,\qquad \tilde f_h(\kk) =
\frac{Z_{h-1}}{\tilde Z_{h-1}(\kk)}f_h(\kk)\;.\label{3.16} \eea
At this point, we can integrate the scale $h$ and, defining
\be e^{\VV^{(h-1)}(A,\sqrt{Z_{h-1}}\psi) + |\L| \tilde
E_{h}} := \int P(d\psi^{(h)}) P(dA^{(h)})e^{\hat \VV ^{(h)}
(A +
A^{(h)} ,\sqrt{Z_{h-1}}(\psi + \psi^{(h)}))}\;,\label{3.17} \ee
our inductive assumption (\ref{3.3}) is reproduced at the scale
$h-1$ with $E_{h-1} := E_{h} + t_{h} + \tilde E_{h}$. Notice that
(\ref{3.17}) can be seen as a recursion relation for the effective 
potential, since from (\ref{3.10}), (\ref{3.13}) it follows that
\be
\hat \VV^{(h)}(A,\sqrt{Z_{h-1}}\psi) = 
\widetilde \VV^{(h)}(A,\sqrt{Z_h}\psi) = \VV^{(h)}(A,\sqrt{Z_h}\psi) - 
\LL_{\psi}\VV^{(h)}(A,\sqrt{Z_h}\psi)\;.\label{3.17b}
\ee
The integration in (\ref{3.17}) is performed by expanding in series
the exponential in the r.h.s. (which involves interactions of any
order in $\psi$ and $A$, as apparent from (\ref{3.5})), and
integrating term by term with respect to the gaussian integration
$P(d\psi^{(h)}) P(dA^{(h)})$. This procedure gives rise to an
expansion for the effective potentials $\VV^{(h)}$ (and to an
analogous expansion for the correlations) in terms of the
renormalized parameters
$\{e_{\m,k},\n_{\m,k},Z_{k-1},v_{k-1}\}_{h< k\le 0}$, which can be
conveniently represented as a sum over {\em Feynman graphs}
according to rules that will be explained below. We will call
$\{e_{\m,k},\n_{\m,k}\}_{h< k\le 0}$ {\it effective couplings} or
{\it running coupling constants} while $\{e_{\m,k}\}_{h< k\le 0}$
are the {\it effective charges}

Note that such {\it renormalized} expansion is significantly
different from the power series expansion in the bare couplings
$e,\n_\m$; while the latter is plagued by {\it logarithmic
divergences}, the former is {\it order by order finite}.

By comparing (\ref{1.1}) and (\ref{1.2}) with (\ref{3.3}),
(\ref{3.5}) and (\ref{3.14}), we see that the integration of the
fields living on momentum scales $\ge M^h$ produces an {\it
effective theory} very similar to the original one, modulo the
presence of a new propagator, involving a renormalized velocity
$v_h$ and a renormalized wave function $Z_h$, and the presence of
a modified interaction $\VV^{(h)}$. The lack of Lorentz symmetry
in our model (implied by the fact that $v\neq 1$) has two main
effects: (1) the Fermi velocity has a non trivial flow; (2) the
marginal terms in the effective potential are defined in terms of
{\it two} charges, namely $e_{0,h}$ and $e_{1,h}=e_{2,h}$, which
are {\it different}, in general.

\subsection{Tree expansion}\label{sec4}

\begin{figure}[htbp]
\centering
\includegraphics[width=0.95\textwidth]{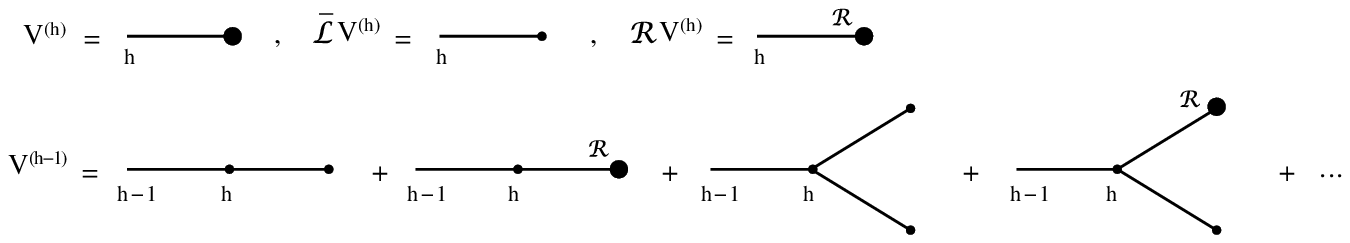} \caption{Graphical
interpretation of Eq.(\protect\ref{4.1}). The graphical equations
for $\lis\LL \VV^{(h-1)}$, $\RR \VV^{(h-1)}$ are obtained from the
equation in the second line by putting an $\lis\LL$, $\RR$ label,
respectively, over the vertices on scale $h$.} \label{fig4.2}
\end{figure}

\begin{figure}[htbp]
\centering
\includegraphics{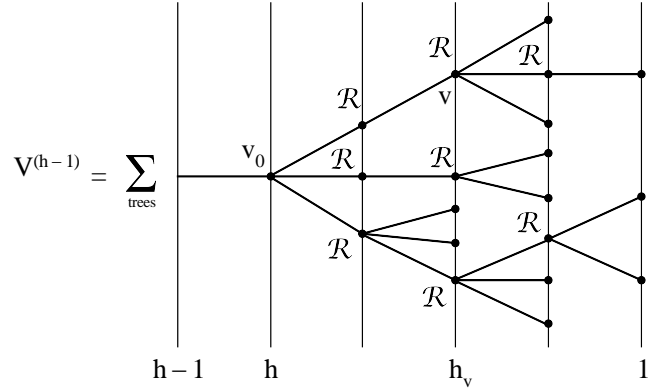} \caption{The effective potential $\VV^{(h-1)}$
can be represented as a sum over {\em Gallavotti -- Nicol\`o}
trees. The black dots will be called {\it vertices} of the tree.
All the vertices except the first ({\it i.e.} the one on scale
$h$) have an $\RR$ label attached, which
 means that they correspond to the action of $\RR\EE^{T}_{h_v}$, while the
first represents $\EE^{T}_{h}$. The endpoints correspond to the
graph elements in Fig.\protect\ref{fig.3b} associated to the two
terms in (\protect\ref{3.14}).} \label{fig:3a}
\end{figure}

The iterative integration procedure described above leads to a representation of 
the effective potentials in terms of a sum over connected Feynman diagrams, 
as explained in the following.
The key formula, which we start from, is (\ref{3.17}), which can
be rewritten as
\bea &&|\L|\tilde E_{h} + \VV^{(h-1)}(A^{(\le
h-1)},\sqrt{Z_{h-1}}\, \psi^{(\le h-1)})=\nn\\&&\quad= \sum_{n\geq
1}\frac{1}{n!}\EE_{h}^{T}\left(\hat \VV^{(h)} (A^{(\le
h)},\sqrt{Z_{h-1}}\psi^{(\le h)});n\right)\;,\label{4.1}\qquad\eea
with $\EE_h^T$ the truncated expectation on scale $h$, defined as
\be \EE_{h}^{T}(X(A^{(h)},\psi^{(h)});n) :=
\frac{\partial^{n}}{\partial \lambda^{n}}\log \int P(d\psi^{(h)})
P(dA^{(h)}) e^{\lambda
X(A^{(h)},\psi^{(h)})}\Big|_{\lambda=0}\;\label{4.2}\ee
If $X$ is graphically represented as a vertex with external lines
$A^{(h)}$ and $\psi^{(h)}$, the truncated expectation (\ref{4.2})
can be represented as the sum over the Feynman diagrams obtained
by contracting in all possible connected ways the lines exiting
from $n$ vertices of type $X$. Every contraction corresponds to a
propagator on scale $h$, as defined in (\ref{3.16}). Since
$\hat{\VV}^{(h)}$ is related to $\VV^{(h)}$ by a rescaling and a
subtraction, see (\ref{3.10}) and (\ref{3.13}), Eq.(\ref{4.1}) can
be iterated until scale $0$, and $\VV^{(h-1)}$ can be written as a
sum over connected Feynman diagrams with lines on all possible
scales between $h$ and $0$. The iteration of (\ref{4.1}) induces a
natural hierarchical organization of the scale labels of every
Feynman diagram, which will be conveniently represented in terms
of tree diagrams. In fact, let us rewrite $\hat \VV^{(h)}$ in the
r.h.s. of (\ref{4.1}) as $\hat
\VV^{(h)}(A,\sqrt{Z_{h-1}}\psi)=\lis\LL\VV^{(h)}
(A,\sqrt{Z_{h}}\psi)+\RR\VV^{(h)}(A,\sqrt{Z_{h}}\psi)$, where
$\lis\LL:=\LL-\LL_{\psi}$, see (\ref{3.10}). Let us graphically
represent $\VV^{(h)}$, $\lis\LL\VV^{(h)}$ and $\RR \VV^{(h)}$ as
in the first line of Fig.\ref{fig4.2}, and let us represent
Eq.(\ref{4.1}) as in the second line of Fig.\ref{fig4.2}; in the
second line, the node on scale $h$ represents the action of
$\EE^T_h$.
Iterating the graphical equation in Fig.\ref{fig4.2} up to scale
0, we end up with a representation of $\VV^{(h)}$ in terms of a
sum over {\it Gallavotti-Nicol\`o} trees $\t$ \cite{G84,BG,GeM}:
\be \VV^{(h)}(A^{(\le h)}, \sqrt{Z_h}\,\psi^{(\le h)}) =
\sum_{N\ge
1}\sum_{\tau\in\mathcal{T}_{h,N}}\VV^{(h)}(\t)\;,\label{4.1a}\ee
where $\TT_{h,N}$ is the set of rooted trees with {\em root} $r$ on
scale $h_{r}=h$ and $N$ endpoints, see Fig.\ref{fig:3a}.
The tree value $\VV^{(h)}(\t)$ can be evaluated in terms of a sum
over connected Feynman diagrams, defined by the following rules.

\begin{figure}[htbp]
\centering
\includegraphics[width=.5\textwidth]{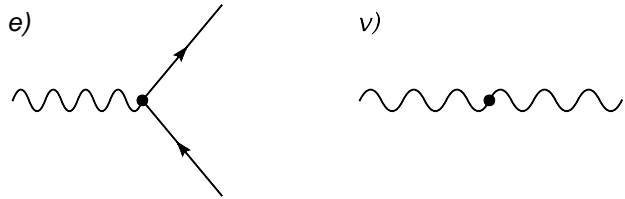}
\caption{The two possible graph elements associated to the
endpoints of a tree, corresponding to the two terms in the r.h.s.
of (\protect\ref{3.14}).} \label{fig.3b}
\end{figure}

With each endpoint $v$ of $\t$ we associate a graph element of
type $e$ or $\n$, corresponding to the two terms in the r.h.s. of
(\ref{3.14}), see Fig. \ref{fig.3b}.
We introduce a {\it field label} $f$ to distinguish the fields
associated to the graph elements $e$ and $\n$ (any field label can
be either of type $A$ or of type $\psi$); the set of field labels
associated with the endpoint $v$ will be called $I_v$.
Analogously, if $v$ is not an endpoint, we call $I_v$ the set of
field labels associated with the endpoints following the vertex
$v$ on $\t$.

\begin{figure}[hbtp]
\centering
\includegraphics[width=.8\textwidth]{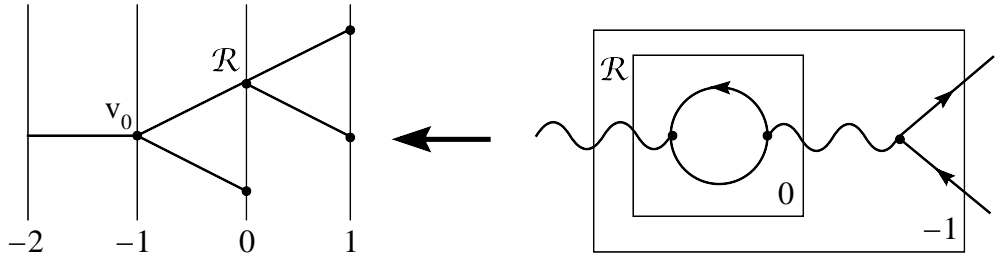}
\caption{A possible Feynman diagram contributing to $V^{(-2)}$ and
its cluster structure.}\label{fig:1e}
\end{figure}

We start by looking at the graph elements corresponding to
endpoints of scale $1$: we group them in {\it clusters}, each
cluster $G_v$ being the set of endpoints attached to the same
vertex $v$ of scale 0, to be graphically represented by a box
enclosing its elements. For any $G_v$ of scale 0 (associated to a
vertex $v$ of scale $0$ that is not an endpoint), we contract in
pairs some of the fields in $\cup_{w\in G_v}I_w$, in such a way
that after the contraction the elements of $G_v$ are connected;
each contraction produces a propagator $g^{(0)}$ or $w^{(0)}$,
depending on whether the two fields are of type $\psi$ or of type
$A$. We denote by $\III_v$ the set of contracted fields inside the
box $G_v$ and by $P_v=I_v\setminus \III_v$ the set of external
fields of $G_v$; if $v$ is not the vertex immediatly following the
root we attach a label $\RR$ over the box $G_v$, which means that
the $\RR$ operator, defined after (\ref{3.5}), acts on the value
of the graph contained in $G_v$. Next, we group together the
scale-0 clusters into scale-(-1) clusters, each scale-(-1) cluster
$G_v$ being a set of scale-0 clusters attached to the same vertex
$v$ of scale $-1$, to be graphically represented by a box
enclosing its elements, see Fig.\ref{fig:1e}.

Again, for each $v$ of scale $-1$ that is not an endpoint, if we
denote by $v_1,\ldots,v_{s_v}$ the vertices immediately following
$v$ on $\t$, we contract some of the fields of
$\cup_{i=1}^{s_v}P_{v_i}$ in pairs, in such a way that after the
contraction the boxes associated to the scale-0 clusters contained
in $G_v$ are connected; each contraction produces a propagator
$g^{(-1)}$ or $w^{(-1)}$. We denote by $\III_v$ the set of fields
in $\cup_{i=1}^{s_v}P_{v_i}$ contracted at this second step and by
$P_v=\cup_{i=1}^{s_v}P_{v_i}\setminus\III_v$ the set of fields
external to $G_v$; if $v$ is not the vertex immediatly following
the root we attach a label $\RR$ over the box $G_v$.

Now, we iterate the construction, producing a sequence of boxes
into boxes, hierarchically arranged with the same partial ordering
as the tree $\t$. Each box $G_v$ is associated to many different
Feynman (sub-)diagrams, constructed by contracting in pairs some
of the lines external to $G_{v_i}$, with $v_i$, $i=1,\ldots,s_v$,
the vertices immediately following $v$ on $\t$; the contractions
are made in such a way that the clusters
$G_{v_1},\ldots,G_{v_{s_v}}$ are connected through propagators of
scale $h_v$. We denote by $P_v^A$ and by $P_v^\psi$ the set of
fields of type $A$ and $\psi$, respectively, external to $G_v$.
The set of connected Feynman diagrams compatible with this
hierarchical cluster structure will be denoted by $\G(\t)$. Given
these definitions, we can write:
\bea &&\VV^{(h)}(\t)=\sum_{\GG\in\G(\t)}\int \prod_{f\in P_{v_0}^\psi}
\frac{d\kk_f}{(2\p)^3}\,\prod_{f\in P_{v_0}^A}\frac{d\pp_f}{(2\p)^3}
\Val(\GG)\;,\nn\\
&&\Val(\GG)=\Big[ \prod_{f\in P_{v_0}^A}A_{\m(f),\,\pp_f}^{(\le
h)}\Big] \Big[\prod_{f\in P_{v_0}^\psi}
\sqrt{Z_{h-1}}\,\widetilde\psi_{\kk_f,\r(f)}^{(\le h)}\Big]\d(v_0)\widehat \Val(\GG)\;,\label{4.4}\\
&&\widehat \Val(\GG)=(-1)^\p\int\!\!\!\prod_{v\ {\rm not}\ {\rm e.p.}}
\Big(\frac{Z_{h_v-1}}{Z_{h_v-2}}\Big)^{\frac{|
P_v^\psi|}2}\frac{\RR^{\a_v}}{s_{v}!} \Biggl[\Big(\prod_{\ell\in
v}g_\ell^{(h_v)}\Big) \Big(\prod_{\substack{v^*\ {\rm e.p.}\\
v^*>v,\\h_{v^*}=h_v+1}} K_{v^*}^{(h_v)}\Big)\Biggr]\nn \eea
where: $(-1)^\p$ is the sign of the permutation necessary to bring
the contracted fermionic fields next to each other; in the product
over $f\in P_v^\psi$, $\widetilde\psi$ can be either $\lis\psi$ or
$\psi$, depending on the specific field label $f$; 
$\d(v_0)=\d\Big(\sum_{f\in P_{v_0}^A}\pp_f-\sum_{f\in P_{v_0}^{\psi}}
(-1)^{\e(f)}\kk_f\Big)$, where $\e(f)=\pm$ depending on whether
$\widetilde\psi$ is equal to $\lis\psi$ or $\psi$; the integral in the third
line runs over the independent loop momenta; $s_v$ is the
number of vertices immediately following $v$ on $\t$; $\RR=1-\LL$
is the operator defined in (\ref{3.6}) and preceding lines);
$\a_v=0$ if $v=v_0$, and otherwise $\a_v=1$; $g_\ell^{(k)}$ is
equal to $g^{(k)}$ or to $w^{(k)}$ depending on the fermionic or
bosonic nature of the line $\ell$, and $\ell\in v$ means that
$\ell$ is contained in the box $G_v$ but not in any other smaller
box; finally, $K_{v^*}^{(k)}$ is the matrix associated to the
endpoints $v^*$ on scale $k+1$ (given by $ie_{0,k}\g_0$ if $v^*$
is of type (a) with label $\r=0$, by $ie_{j,k}v_k\g_j$ if $v^*$ is
of type $e$ with label $\r=j\in\{1,2\}$, or by $-M^k\n_{\m,k}$ if $v^*$ is
of type $\n$. In (\ref{4.4}) it is understood that the operators
$\RR$ act in the order induced by the tree ordering (i.e.,
starting from the endpoints and moving toward the root); moreover,
the matrix structure of $g^{(k)}_\ell$ is neglected, for
simplicity of notations. 

\subsection{An example of Feynman graph}

\begin{figure}[hbtp]
\centering
\includegraphics[height=3.5truecm]{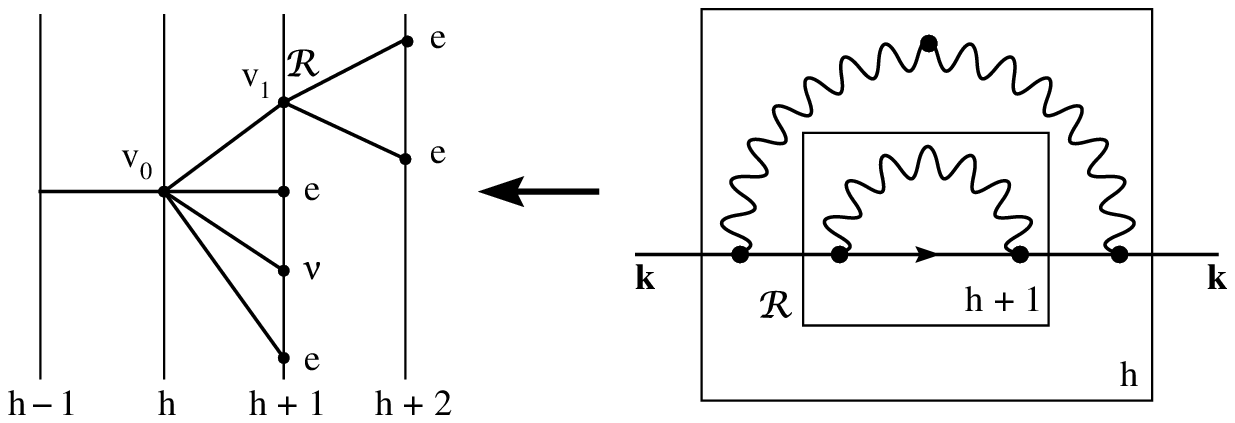}
\caption{A possible Feynman diagram contributing to $\VV^{(h-1)}$
and its cluster structure.}\label{figex}
\end{figure}

To be concrete, let us apply the rules described above in the
evaluation of a simple Feynman graph $\GG$ arising in the tree
expansion of $\VV^{(h-1)}$. Let $\GG$ be the diagram in
Fig.\ref{figex}, associated to the tree $\t$ drawn in the left
part of the figure; let us assume that the sets $P_v$ of the
external lines associated to the vertices of $\t$ are all
assigned.
Setting
\be e_{0,h} := e_{0,h}\,,\qquad \bar e_{j,h} := v_{h-1} e_{j,h}\;,
\ee
we can write:
\bea &&\Val(\GG) = -\frac{1}{4!2!}\frac{Z_h}{Z_{h-1}}
\frac{Z_{h-1}}{Z_{h-2}}\bar e_{\m_1,h}^2\bar
e_{\m_2,h+1}^{2}M^{h}\n_{\m_1,h}\lis\psi_{\kk}\cdot\nn\\&&\quad\cdot\Biggl\{\int
\frac{d\pp}{(2\pi)^3}\,|w^{(h)}(\pp)|^2 \g_{\m_1}
g^{(h)}(\kk+\pp)\cdot\label{4.6}\\&& \cdot\RR\Big[ \int
\frac{d\qq}{(2\pi)^3}\, \g_{\m_2}g^{(h+1)}(\kk+\pp+\qq)\g_{\m_2}
w^{(h+1)}(\qq) \Big]
g^{(h)}(\kk+\pp)\g_{\m_1}\Biggr\}\psi_\kk\;,\nn\eea
where $\RR \big[F(\kk+\pp)\big] = F(\kk+\pp)-F(\V0)-(\kk+\pp)\cdot
\nabla F(\V0)\equiv \frac12(k_\m+p_\m)(k_\n+p_\n)\dpr_\m\dpr_\n
F(\kk^*)$. Notice that the same
 Feynman graph appears in the evaluation of other trees, which are
topologically equivalent to the one represented in the left part
of Fig.\ref{figex} and that can be obtained from it by: (i)
relabeling the fields in $P_{v_1}$, $P_{v_0}$, (ii) relabeling the
endpoints of the tree, (iii) exchanging the relative positions of
the topologically different subtrees with root $v_0$. If one sums
over all these trees, the resulting value one obtains is the one
in Eq.(\ref{4.6}) times a combinatorial factor $2^2\cdot 3\cdot 4$
($2^2$ is the number of ways for choosing the fields in $P_{v_1}$
and in $P_{v_0}$; $3$ is the number of ways in which one can
associate the label $\n$ to one of the endpoints of scale $h+1$;
$4$ is the number of distinct unlabelled trees that can be
obtained by exchanging the positions of the subtrees with root
$v_0$). 

\subsection{Dimensional bounds}\label{bound} We are now ready to derive
a general bound for the Feynman graphs produced by the multiscale
integration. Let $W^{N;(h)}_{m,n,\ul{\r},\ul{\m}}$ be the
contribution from trees with $N$ end-points to the kernel
$W^{(h)}_{m,n,\ul{\r},\ul{\m}}$ in \ref{3.5}, that is
\be W^{(h)}_{m,n,\ul{\r},\ul{\m}}(\{\kk_i\},\{\pp_j\})=\sum_{N=1}^\io
\sum_{\t\in \TT_{h,N}}\sum^*_{\substack{\GG\in\G(\t)\\
|P^A_{v_0}| = m,\\|P^{\psi}_{v_0}| = 2n}} \widehat\Val(\GG)\equiv
\sum_{N=1}^\io W^{N;(h)}_{m,n,\ul{\r},\ul{\m}}(\{\kk_i\},\{\pp_j\})\;,\label{WN}\ee
where the * on the sum indicates the constraints that: 
$\cup_{f\in P_{v_0}^A}\{\pp_f\}=\cup_{j=1}^m\{\pp_j\}$; $\cup_{f\in P_{v_0}^\psi}\{\kk_f\}=
\cup_{i=1}^{2n}\{\kk_i\}$; $\cup_{f\in P_{v_0}^A}\{\m(f)\}=\underline\m$; 
$\cup_{f\in P_{v_0}^\psi}\{\r(f)\}=\underline \r$.

The $N$-th order contribution to the kernel of the effective potential 
admits the following bound. 
\begin{theorem}{\bf{($N!$ bound)}}\label{thm1} Let $\bar\e_h =
\max_{h< k\leq 0} \{|e_{\m,k}|,|\n_{\m,k}|\}$ be small enough. If
$Z_{k}/Z_{k-1} \le e^{C\bar \e_h^2}$ and ${C}^{-1}\le v_{k-1}\le 1$,
for all $h< k\le 0$ and a suitable constant $C>0$, then
\be ||W^{N;(h)}_{m,n,\ul{\r},\ul{\m}}|| \le (\const.)^N\bar\e_h^N
\Big(\frac{N}2\Big)!\; M^{h(3-m-2n)}\;,\label{4.11a}\ee
where $||W^{N;(h)}_{m,n,\ul{\r},\ul{\m}}||:=\sup_{\{\kk_i\},\{\pp_j\}}
|W^{N;(h)}_{m,n,\ul{\r},\ul{\m}}(\{\kk_i\},\{\pp_j\})|$.
\end{theorem}
\vskip.3cm The factor $3-2n-m$ in \pref{4.11a} is referred to as
the {\it scaling dimension} of the kernel with $2n$ external
fermionic fields and $m$ external bosonic fields; according to the
usual RG therminology, kernels with positive, vanishing or
negative scaling dimensions are called {\it relevant, marginal} or
{\it irrelevant} operators, respectively. Notice that, if we tried
to expand the effective potential in terms of the bare couplings
$e,\n_\m$, the $N$-th order contributions in this ``naive''
perturbation series could {\it not be bounded uniformly in the
scale $h$} as in (\ref{4.11a}), but rather by the r.h.s. of
(\ref{4.11a}) times $|h|^N$, an estimate which blows up order by
order as $h\to-\io$. \vskip.3cm

\noindent{\it Proof.}
Using the bounds
\bea&&\big\| g^{(h)}(\kk) \big\| \leq \const\cdot
M^{-h}\;,\,\qquad
\int d\kk \big\| g^{(h)}(\kk) \big\| \leq \const\cdot M^{2h}\;,\nn\\
&&\big| w^{(h)}(\kk) \big| \leq \const\cdot M^{-h}\;,\,\qquad \int
d\kk \big|w^{(h)}(\kk) \big| \leq\const\cdot M^{2h}\;,\label{4.7}
\eea
and the assumptions on $v_{k-1}$ and $Z_k/Z_{k-1}$ 
into (\ref{4.4}), we find that, if $\t\in\TT_{h,N}$ and
$\GG\in\G(\t)$,
\bea &&|\widehat\Val(\GG)|\le (\const.)^N\bar\e_h^N\cdot\label{4.8}\\
&&\hskip.5cm\cdot\prod_{v\ {\rm not}\ {\rm e.p.}} \frac{e^{\frac{C}2\bar \e_h^2
|P_v^\psi|}}
{s_v!}M^{-3 h_v(s_v-1)}M^{2 h_v n^0_v} M^{h_v m^\n_v} \prod_{\substack{v\ {\rm
not}\ {\rm e.p.}\\ v>v_0}} M^{-z_v(h_v-h_{v'})}\;,\nn\eea
where: $n^0_v$ is the number of propagators $\ell\in v$, i.e., of
propagators $\ell$ contained in the box $G_v$ but not in any
smaller cluster; $s_v$ is the number of vertices immediately
following $v$ on $\t$; $m^\n_v$ is the number of end-points of
type $\n$ immediately following $v$ on $\t$ (i.e., contained in
$G_v$ but not in any smaller cluster); $v'$ is the vertex
immediately preceding $v$ on $\t$ and $z_v=2$ if
$|P_v^\psi|=|P_v|=2$, $z_v=1$ is $|P_v^\psi|=2|P_v^A|=2$ and
$z_v=0$ otherwise. The last product in (\ref{4.8}) is due to the
action of $\RR$ on the vertices $v>v_0$ that are not end-points.
In fact, the operator $\RR$, when acting on a kernel
$W^{(h_v)}_{1,1}(\pp,\kk)$ associated to a vertex $v$ with
$|P_v^\psi|=2|P_v^A|=2$, extracts from $W^{(h_v)}_{1,1}$ the rest
of first order in its Taylor expansion around $\pp=\kk=\V0$: if
$|W^{(h_v)}_{1,1}(\pp,\kk)|\le C(v)$, then $|\RR
W^{(h_v)}_{1,1}(\pp,\kk)|=\frac12|(\pp\dpr_\pp+\kk\dpr_\kk)
W^{(h_v)}_{1,1}(\pp^*,\kk^*)|\le (\const.) M^{-h_v+h_{v'}}C(v)$,
where $M^{-h_{v}}$ is a bound for the derivative with respect to
momenta on scale $h_v$ and $M^{h_{v'}}$ is a bound for the
external momenta $\pp$, $\kk$; i.e., $\RR$ is dimensionally
equivalent to $M^{-(h_v-h_{v'})}$. The same is true if $\RR$ acts on 
a kernel $W^{(h_v)}_{3,0}(\pp_1,\pp_2)$. Similarly, if $\RR$ acts on a
terms with $|P_v|=2$, it extracts the rest of second
order in the Taylor expansion around $\kk=\V0$, and it is
dimensionally equivalent to $\kk^2\dpr_\kk^2\sim
M^{-2(h_v-h_{v'})}$. As a result, we get (\ref{4.8}).

Now, let $n^e_v$ ($n^\n_v$) be the number of vertices of type $e$
(of type $\n$) following $v$ on $\t$. If we plug in (\ref{4.8})
the identities
\bea &&\sum_{v\ {\rm not}\ {\rm e.p.}} (h_v-h)(s_v-1) =
\sum_{v\ {\rm not}\ {\rm e.p.}} (h_v-h_{v'})(n^e_v+n^\n_v-1)\nn\\
&&\sum_{v\ {\rm not}\ {\rm e.p.}}(h_v - h)n^{0}_v = \sum_{v\ {\rm
not}\ {\rm e.p.}}(h_v - h_{v'})\Big( \frac{3}{2}n^{e}_v +
n^{\n}_{v} -
\frac{|P_{v}|}{2} \Big)\nn\\
&&\sum_{v\ {\rm not}\ {\rm e.p.}} (h_v - h)m^{\n}_{v} = \sum_{v\
{\rm not} \ {\rm e.p.}}(h_{v} - h_{v'})n^{\n}_{v}\label{4.9} \eea
we get the bound
\be|\widehat\Val(\GG)|\le (\const.)^N{\bar\e}_h^N\frac1{s_{v_0}!}
M^{h(3-|P_{v_0}|)}\!\!\!\prod_{ \substack{v\ {\rm not}\ {\rm e.p.}\\
v>v_0}}\frac{e^{\frac{C}2\bar \e_h^2 
|P_v^\psi|}}{s_v!}M^{(h_v - h_{v'}) (3-|P_v|-z_v)}\;.\label{4.10}\ee
In the latter equation, $3-|P_v|$ is the {\it scaling dimension}
of the cluster $G_v$, and $3-|P_v|-z_v$ is its renormalized
scaling dimension. Notice that the renormalization operator $\RR$
has been introduced precisely to guarantee that $3-|P_v|-z_v<0$
for all $v$, by construction. This fact allows us to sum over the
scale labels $h\le h_v\le 1$, and to conclude that the
perturbative expansion is well defined at any order $N$ of the
renormalized expansion. More precisely, the fact that the
renormalized scaling dimensions are all negative implies, via a
standard argument (see, e.g., \cite{BG,GeM}), the following
bound, valid for a suitable constant $C$ (see (\ref{4.11a})
for a definition of the norm $\|\cdot\|$):
\bea &&||W^{N;(h)}_{m,n,\ul{\r},\ul{\m}}|| \le (\const.)^N\bar\e_h^N 
\frac1{s_{v_0}!}M^{h(3-m-2n)}\cdot\label{A.11}\\&&
\hskip2.truecm\cdot\sum_{\t\in \TT_{h,N}}\sum_{\substack{\GG\in\G(\t)\\
|P^A_{v_0}| = m,\\|P^{\psi}_{v_0}| = 2n}}\prod_{\substack{v\ {\rm
not}\ {\rm e.p.}\\ v>v_0}}\frac{e^{\frac{C}2\bar \e_h^2
|P_v^\psi|}}{s_v!}M^{(h_v - h_{v'})(3 - |P_v| - z_v)}\;,
\nn\eea
from which, after counting the number of Feynman graphs
contributing to the sum in (\ref{A.11}), (\ref{4.11a})
follows.\qed

An immediate corollary of the proof leading to (\ref{4.11a}) is
that contributions from trees $\t\in\TT_{h,N}$ with a vertex $v$ on
scale $h_v=k>h$ admit an improved bound with respect to
(\ref{4.11a}), of the form $\le (\const.)^N\bar\e_h^N (N/2)!\,
M^{h(3-|P_{v_0}|)}M^{\th(h-k)}$, for any $0<\th<1$; the factor
$M^{\th(h-k)}$ can be thought of as a dimensional gain with
respect to the ``basic'' dimensional bound in (\ref{4.11a}). This
improved bound is usually referred to as the {\it short memory}
property (i.e., long trees are exponentially suppressed); it is due 
to the fact that the renormalized scaling dimensions $d_v=3-|P_v|-z_v$
in (\ref{4.10})  are all negative, and can be obtained by taking a fraction of 
the factors $M^{(h_v-h_{v'})d_v}$ associated to the branches of the tree
$\t$ on the path connecting the vertex on scale $k$ to the one on scale $h$.

{\bf Remark.} All the analysis above is based on the fact that the
scaling dimension $3-|P_v|$ in (\ref{4.10}) is independent of the
number of endpoints of the tree $\t$; i.e., the model is {\it
renormalizable}. A rather different situation is found in the case
of instantaneous Coulomb interactions, in which case the bosonic
propagator is given by $(2|\vec p|)^{-1}$ rather than by
$(2|\pp|)^{-1}$. In this case, choosing the bosonic single scale
propagator as $w^{(h)}(\pp)=\c_0(\pp)f_h(\vec p)(2|\vec p|)^{-1}$,
one finds that the last bound in (\ref{4.7}) is replaced by $\int
d\pp \big|w^{(h)}(\pp) \big|\le(\const.)\, M^{h}$ (dimensionally,
this bound has a factor $M^h$ missing). Repeating the steps
leading to (\ref{4.10}), one finds a general bound valid at all
orders, in which the new scaling dimension is
$3-|P_v|+n^e_v+n^\n_v$; this (pessimistic) general bound assumes
that at each scale the loop lines of the graph are all bosonic.
Perhaps, this bound can be improved, by taking into account the
explicit structure of the expansion; however, it shows that the
renormalizability of the instantaneous case, {\it if true}, does
not follow from purely dimensional considerations and its proof
will require the implementation of suitable cancellations.

\begin{table}
\begin{center}
    \begin{tabular}{|l|p{13cm}|}
  \hline
    Symbol & Description \\ \hline
    $\t$ & Gallavotti -- Nicol\` o (GN) tree.\\
    $r$ & Root label of the tree.\\
    $v_0$ & First vertex of the tree, immediately following the root.\\
    $h_v$ & Scale label of the tree vertex $v$.\\
    $\TT_{h,N}$ & Set of GN trees with root on scale $h_r = h$ 
    and with $N$ endpoints.\\
    $I_v$ & Set of field labels associated with the endpoint of the tree $v$.\\
    $G_v$ & Cluster associated with the tree vertex $v$.\\
    $\III_v$ & Set of contracted fields inside the box corresponding to the cluster $G_v$.\\
    $P_v$ & Set of external fields of $G_v$.\\
    $v_{i}$ & $i$-th vertex immediately following $v$ on the tree.\\
    $s_{v}$ & Number of vertices immediately following the vertex $v$ on the tree.\\
    $P_{v}^{\#}$ & Set of fields of type $\# = A,\psi$ external to $G_v$.\\
    $\G(\t)$ & Set of connected Feynman diagrams compatible 
    with the hierarchical cluster structure of the tree $\t$.\\
    $n^{0}_{v}$ & Number of propagators contained in $G_{v}$ but not in any smaller cluster.\\
    $m^{\n}_{v}$ & Number of end-points of type $\n$ immediately following $v$ on the tree.\\
    $v'$ & Vertex immediately preceding $v$ on the tree.\\
    $n^{\#}_{v}$ & Number of vertices of type $\# = e,\n$ following $v$ on the tree.\\
    $z_v$ & Improvement on the scaling dimension due to the renormalization.\\
    \hline
    \end{tabular}
\end{center}
\caption{List of the symbols introduced in Sections \ref{sec4}, \ref{bound}.}
\end{table}

\subsection{The Schwinger functions}\label{schwing}

A similar analysis can be performed for the 2-point function, see Appendix \ref{app3}.
It turns out that, similarly to what we found above for the effective potentials,
the 2-point function can be written in terms of a 
renormalized perturbative expansion in the effective couplings
$\{e_{\m,k},\n_{\m,k}\}_{k\leq 0}$ and in the renormalization constants 
$\{Z_k,v_{k}\}_{k\leq 0}$, with coefficients represented as sums of Feynman
graphs, uniformly bounded as $|\L| \to \io$; in contrast, the
graphs forming the naive expansion in $e,\n_{\m}$ are plagued by
logarithmic infrared divergences.

More explicitly, if  $M^{h}\leq |\kk|\leq M^{h+1}$, we get (see Eqs.(\ref{A3.21})--(\ref{A3.220})):
\be \media{\psi_\kk\lis\psi_\kk} = \sum_{j=h}^{h+1}
\frac{g^{(j)}(\kk)}{Z_{j-1}}\Big(1 + \tilde B(\kk)\Big)\;,\label{4.12}\ee
where $\tilde B(\kk)$ is given by a formal power series in
$\{e_{\m,k},\n_{\m,k}\}_{k\leq 0}$ with coefficients depending on
$\{Z_k,v_k\}_{k\le 0}$, and starting from second order; under the
same hypothesis of Theorem \ref{thm1}, the $N$-th order
contribution to $\tilde B(\kk)$ is bounded by
$(\const.)^{N}(\bar\e_{-\io})^N (N/2)!$ uniformly in $\kk$. Eq.(\ref{4.12})
is equivalent to Eq.(\ref{1.7}) of the main result (see the end of Section
\ref{secflowZv} below for the explicit relation between 
$Z_h,v_h$ and $Z(\kk),v(\kk)$).

To prove our main result we need to control the flow of the
effective charges at all orders in perturbation theory, and to do
this we shall use Ward Identities, see Section \ref{sec5}. These
are nontrivial relations for the three point functions, which can
be related to the renormalized charges in the following way.
Consider a theory with a bosonic infrared cutoff $M^{h_{*}}$, that
is assume that the bare bosonic propagator is given by (\ref{1.5})
with $\chi_0(\pp)$ replaced by $\chi_{[h^{*},0]}(\pp):=
\chi_0(\pp)-\chi_0(M^{-h^{*}}\pp)$, which is vanishing for
$|\pp|\leq M^{h_*}$ and it is equal to $\chi_0(\pp)$ for
$|\pp|\geq M^{h^{*}+1}$; denote by $\media{\ldots}_{h^*}$ the
expectation value in the presence of the bosonic infrared cutoff.
As shown in Appendix \ref{app3}, 
setting $\bar e_{0,h}:= e_{0,h}$,
$\bar e_{1,h}= \bar e_{2,h}:= v_{h-1} e_{1,h}$, and taking
$|\qq| = M^{h^{*}}$, $|\qq+\pp|\leq M^{h^{*}}$, $|\pp|\ll M^{h^{*}}$ 
(we will be interested in the limit $\pp\rightarrow \V0$), the
following result holds (see Eqs.(\ref{A3.24})--(\ref{A3.2302})):
\be
\media{j_{\m,\pp};\psi_{\qq+\pp}\lis\psi_{\qq}}_{h^{*}} =
iZ_{h^{*}-1}
\frac{\bar e_{\m,h^{*}}}
{e}\media{\psi_{\qq+\pp}\lis\psi_{\qq+\pp}}_{h^{*}}
\Big(\g_\m +
\bar B_{\m,h^{*}}(\pp,\qq)\Big)\media{\psi_{\qq}\lis\psi_{\qq}}_{h^{*}}\;,
\label{4.12a}
\ee
where $\bar B_{\m,h^{*}}$ is given by a formal power series in
$\{e_{\m,k},\n_{\m,k}\}_{h^{*}<k\le 0}$, starting from second order and 
with the $N$-th order of the series
admitting a bound proportional to $(\bar\e_{h^*})^N(N/2)!$, uniformly in $\kk$.
Eq.(\ref{4.12a}) is one of the two desired equations relating the 3-point
function to the 2-point function and the effective charge $e_{\m,h^*}$.
A second independent equation expressing the 3-point function in terms of the
2-point function and of the {\it bare} charge $e$ will be derived in Section
\ref{sec5}, see (\ref{5.2}), using the (approximate) gauge invariance of the
theory. Combining the two equations we will be able to relate $e_{\m,h^*}$
to the bare charge $e$, for all $h^*<0$, and this will allows us to control
the flow of the effective couplings on all infrared scales.
This procedure will be described in detail in the next two sections.

\section{The flow of the effective couplings}
\label{sec3a}
\setcounter{equation}{0}
\renewcommand{\theequation}{\ref{sec3a}.\arabic{equation}}
\subsection{The Beta function}
A crucial point for the consistency of our approach is
that the running coupling constants $e_{\m,h},\n_{\m,h}$ are small
for all $h\le 0$, that the ratios $Z_h/Z_{h-1}$ are close to 1, and
the effective Fermi velocity $v_h$ does not approach zero.
Even if we do not prove the convergence of the series but only $N!$ bounds, we
expect that our series gives meaningful information only as long
as the running coupling constants satisfy these conditions.
In this section we describe how to control their flow. We shall proceed
by induction: we will first assume that $\bar\e=\max_{k\le 0}\{|e_{\m,k}|\}$
is small, that $Z_h/Z_{h-1}\le e^{C\bar\e^2}$ and $C^{-1}\le v_h\le 1$ for all
$h\le 0$ and a suitable constant $C>0$, and we will show that,
by properly choosing the values of the counterterms $\n_\m$ in (\ref{1.2}),
the constants $\n_{\m,h}$ remain small: $\max_{h\le 0}\{|\n_{\m,h}|\}\le
(\const.)\,\bar\e^2$. Next, once that the flow of $\n_{\m,h}$ is controlled,
we will study the flow of $Z_h$ and $v_h$ under the assumption that 
the constants $e_{\m,h}$ remain bounded and small for
all $h\le 0$; we will show that,
asymptotically as $h\to-\io$, $Z_h\sim M^{-\h h}$, with $\h=O(e^2)$
a positive exponent, while $v_h$ grows, approaching a limiting value $v_{eff}$
close to the speed of light. Finally, we shall start to discuss the 
remarkable cancellations following from a {\it Ward Identity} that 
guarantee that the constants $e_{\m,h}$ remain bounded and small for
all $h\le 0$; the full proof of this fact will be postponed to Section 
\ref{sec5} and Appendix \ref{app4}.

The renormalized parameters obey to recursive equations induced by
the previous construction; i.e., (\ref{3.7}), (\ref{3.8}),
(\ref{3.12}), (\ref{3.13b}) imply the flow equations:
\bea
&& \frac{Z_{h-1}}{Z_h} = 1 + z_{0,h} := 1 + \b^{z}_{h}\;,\quad v_{h-1} =
\frac{Z_h}{Z_{h-1}}(v_{h} + z_{1,h}) := v_{h} + \b^{v}_{h}\;
\label{3a.18}\\
&&\n_{\m,h} = -M^{-h}\, W^{(h)}_{2,0,\m,\m}(\V0)
:= M \n_{\m,h+1} + \b^{\n}_{\m,h+1}\;,\qquad\label{3a.18n}\\
&&e_{0,h} = \frac{Z_h}{Z_{h-1}}\l_{0,h}:= e_{0,h+1}+\b^{e}_{0,h+1}\;,
\label{3a.18e0}\\
&&e_{1,h} = \frac{Z_h}{Z_{h-1}}\frac{\l_{1,h}}{v_{h-1}}:=
e_{1,h+1}+\b^{e}_{1,h+1}\;,\label{3a.18e1}\eea
and $e_{2,h}=e_{1,h}$. The {\it beta functions} appearing in the
r.h.s. of flow equations are related, see (2.7), to the kernels
$W^{N;(h)}_{m,n,\ul{\r},\ul{\m}}$, so that they are expressed by
series in the running coupling constants admitting the bound
(\ref{4.11a}). For the explicit expressions of the one-loop
contributions to the beta function, see below.

\subsection{The flow of $\n_{\m,h}$}
Let us assume that $\bar\e=\max_{k\le 0} \{|e_{\m,k}|\}$ is small,
that $Z_h/Z_{h-1}\le e^{C\,\bar\e^2}$ and $C^{-1}\le v_h\le 1$ for a suitable 
constant $C$, for all $h\le 0$. Under these assumptions, the
flow of $\n_{\m,h}$ can be controlled by suitably choosing the
counterterms $\n_{\m}$; in fact, if $\n_\m$ is chosen as
\be \n_{\m} =
-\sum_{k=-\io}^{0}M^{k-1}\b^{\n}_{\m,k}\;,\label{4.100}\ee
then the effective coupling $\n_{\m,h}$ is
\be \n_{\m,h} = -\sum_{k=-\io}^{h}M^{-h-1+k}\b^{\n}_{\m,k}\;,
\label{4.101}\ee
from which one finds that $\n_{\m,h}$ can be expressed by a series in 
$\{e_{\m,k}\}_{k\le 0}$, starting at second order and with coefficients bounded
uniformly in $h$. At lowest order, if $h<0$ and setting
$\x_{h}\=\frac{\sqrt{1-v_h^2}}{v_h}$ (see Appendix \ref{app2}):
\bea &&\b^{\n,(2)}_{0,h} = -(M-1)\frac{e_{0,h}^2
v_h^{-2}}{\pi^2}\Big[ \frac{\x_h -
\arctan\x_h}{\x_h^3}\Big]\int_0^\io dt\,\big(2\c(t)-
\c^2(t)\big)\;\label{3a.19n}\\
&&\b^{\n,(2)}_{1,h} = -(M-1)\frac{e_{1,h}^2}{2\pi^2}\Big[
\frac{\arctan\x_h}{\x_h} - \frac{\x_h - \arctan\x_h}{\x_h^3}\Big]
\cdot\nn\\&&\hskip6cm\cdot\int_0^\io
dt\,\big(2\c(t)-\c^2(t)\big)\;.\label{3a.19nn} \eea
By the above equations we see that lowest order contributions to
$\n_\m$ are positive, that is $\n_\m$ can be interpreted as {\it
bare} photon masses. By using the short memory property and
symmetry considerations,
one can also show that $\b^{\n}_{0,h}-\b^{\n}_{1,h}$ is a sum of graphs
whose contributions are of the order $O(1-v_h)$ or $O(e_{0,h}-e_{1,h})$.

\subsection{The flow of $Z_h$ and $v_h$}\label{secflowZv}

In this section we show that, under proper assumptions on the flow of the effective charges,
the effective Fermi velocity $v_h$ tend to a limit value $v_{eff}=v_{-\io}$ and that both 
$v_{eff}-v_h$ and $Z_h^{-1}$ vanish as $h\to-\io$ with an anomalous power law. 

Let us assume that the effective charges
tend to a line of fixed points: 
\be e_{\m,h}=e_{\m,-\io}+O(e^3(v_{-\io}-v_h))+O(e^3M^{\th
h})\;,\label{ch1}\ee
with $0<\th<1$ and $e_{\m,-\io}=e+O(e^3)$; this is a 
remarkable property that will be proven order by order in perturbation theory 
by using WIs, see the following section. Moreover, let $\n_\m$ be fixed as in the 
previous subsection (under the proper inductive assumptions on $Z_k$ and 
$v_k$). 

We start by studying the flow of the Fermi velocity. At lowest order
(see Appendix \ref{app2}), its beta 
function reads:
\bea&&
\b^{v,(2)}_{h} =\label{3a.19v}\\&&=\frac{ \log M}{4\p^2}
\Biggl[\frac{e^2_{0,h}v_h^{-1}}2\frac{\arctan \x_h}{\x_h}- \Big(2
e^2_{1,h}v_{h} - \frac{e^2_{0,h}v_h^{-1}}2\Big)
\frac{\x_h-\arctan\x_h}{\x_h^3}\Biggr]\;.\nn \eea
Note that if $e_{0,k}\=e_{1,k}$, then the r.h.s. of (\ref{3a.19v}) is strictly 
positive for all $\x_h>0$ and it vanishes quadratically in $\x_h$ at 
$\x_h= 0$. 
The higher order contributions to $\b_h^v$ have similar properties. 
This can be proved as follows: we observe that the
beta function $\b^{v}_h$ is a function of the renormalized 
couplings and of the Fermi velocities on scales $\geq h$, {\it i.e.}:
\be
\b^{v}_{h} = \b^{v}_h\big(\big\{(e_{0,k},e_{1,k},e_{2,k}), (\n_{0,k}, \n_{1,k}, \n_{2,k}), v_k
\big\}_{k\ge h}\big)\;.
\label{dec1}
\ee
%
We can rewrite $\b^{v}_h$ as 
$\b^{v,rel}_{h} + \b^{v,1}_h + \b^{v,2}_h + \b^{v,3}_{h}$, with:
\bea
&&\b^{v,rel}_{h} = \b^{v}_h\big(\big\{(e_{0,k},e_{0,k},e_{0,k}), (\n_{0,k}, \n_{0,k}, \n_{0,k}), 1
\big\}_{k\ge h}\big)\;,
\label{dec2}\\
&&\b^{v,1}_{h} \hskip.225truecm = \b^{v}_h\big(\big\{(e_{0,k},e_{0,k},e_{0,k}), (\n_{0,k}, \n_{0,k}, \n_{0,k}), v_k
\big\}_{k\ge h}\big)-\nn\\
&&\hskip1.1truecm -\b^{v}_h\big(\big\{(e_{0,k},e_{0,k},e_{0,k}), (\n_{0,k}, \n_{0,k}, \n_{0,k}), 1
\big\}_{k\ge h}\big)\;,\nn\\
&&\b^{v,2}_{h} \hskip.225truecm= \b^{v}_h\big(\big\{(e_{0,k},e_{0,k},e_{0,k}), (\n_{0,k}, \n_{1,k}, \n_{2,k}), v_k
\big\}_{k\ge h}\big)-\nn\\
&&\hskip1.1truecm - \b^{v}_h\big(\big\{(e_{0,k},e_{0,k},e_{0,k}), (\n_{0,k}, \n_{0,k}, \n_{0,k}), v_k
\big\}_{k\ge h}\big)\;,\nn\\
&&\b^{v,3}_{h} \hskip.225truecm= \b^{v}_h\big(\big\{(e_{0,k},e_{1,k},e_{2,k}), (\n_{0,k}, \n_{1,k}, \n_{2,k}), v_k
\big\}_{k\ge h}\big)-\nn\\
&&\hskip1.1truecm -\b^{v}_h\big(\big\{(e_{0,k},e_{0,k},e_{0,k}), (\n_{0,k}, \n_{1,k}, \n_{2,k}), v_k
\big\}_{k\ge h}\big)\;.\nn
\eea
By relativistic invariance it follows that $\b^{v,rel}_{h}=0$ and by the short 
memory property (see discussion after Eq.(\ref{A.11})) we get:
\be
\b^{v,1}_{h} = O\big(e_{0,h}^{2}(1 - v_{h})\big)\;,\quad \b^{v,2}_{h} = O\big(e^2_{0,h}(\n_{0,h} - \n_{1,h})\big)\;,
\quad \b^{v,3}_{h} = O\big(e_{0,h}(e_{0,h} - e_{1,h})\big)\;.\label{dec3}
\ee
Using (\ref{4.101}) and an argument similar to the one leading to Eq.(\ref{dec3}), 
we also find that 
$\n_{0,h} - \n_{1,h}$ can be written as a sum of contributions
of order $e_{0,h}(e_{0,h} - e_{1,h})$ and of order $e_{0,h}^{2}(1 - v_{h})$. Therefore,  
we can write:
\be \frac{v_{h-1}}{v_h}=1 +\frac{ \log M}{4\p^2} \Biggl[\frac85
e^2(1-v_h)(1+A'_h)+\frac43e(1+B'_h)
(e_{0,h}-e_{1,h})\Biggr]\;,\label{5.11b}\ee
where: the numerical coefficients are obtained from the explicit lowest order computation 
(\ref{3a.19v}); $A'_h$ is a sum of contributions that are finite at all orders
in the effective couplings, which are either of order two or more in the 
effective charges, or vanishing at $v_k=1$; similarly, $B'_h$ is a sum of 
contributions that are finite at all orders in the effective couplings, which 
are of order two or more in the effective charges. From
\pref{5.11b} it is apparent that $v_h$ tends as $h\to-\io$ to a
limit value
\be v_{eff}=1+\frac{5}{6e}(e_{0,-\io}-e_{1,-\io})(1+C'_{-\io}) \label{veff}\ee
with $C'_{-\io}$ a sum of contributions 
that are finite at all orders in the effective couplings, which 
are of order two or more in the effective charges. The fixed point (\ref{veff}) 
is found simply by requiring that in the limit $h\rightarrow-\infty$ 
the argument of the square brakets in (\ref{5.11b}) vanishes. 

Using Eq.(\ref{ch1}), we find that the expression in square 
brackets in the r.h.s. of (\ref{5.11b}) can be rewritten as 
$(8e^2/5)(v_{eff}-v_h+R_h')(1+A''_h)$, where: (i) $A''_h$  
is a sum of contributions that are finite at all orders
in the effective couplings, which are either of order two or more in the 
effective charges, or vanishing at $v_k=v_{eff}$; (ii) $R_h'$ 
is a sum of contributions that are finite at all orders
in the effective couplings, which are of order two or more in the 
effective charges and are bounded at all orders by $M^{\th h}$, for some 
$0<\th<1$. Therefore, (\ref{5.11b})
can be rewritten as
\be
v_{eff} - v_{h-1} = (v_{eff} - v_{h})
\Big(1 - v_{h}\frac{v_{eff} - v_{h}+R_h'}{v_{eff} - v_{h}}
\log M \frac{2 e^{2}}{5\pi^2}(1 + A''_{h}) \Big)\;,\label{veff2}
\ee
from which, using the fact that $R_h'=O(e^2 M^{\th h})$,  
we get that there exist two positive constants $C_1,C_2$
such that \footnote{Eq.(\ref{flowv1}) must be understood as an order by 
order inequality: if we truncate the theory at order $N$ in the bare coupling $e$, 
both sides of the inequality in Eq.(\ref{flowv1}) are verified asymptotically as $e\to 0$, for 
all $N\ge 1$.}:
\be C_1M^{h\tilde\eta}\le \frac{v_{eff} - v_{h}}{v_{eff}-v}\le C_2  M^{h\tilde\eta}\;,
\label{flowv1}\ee
with 
\be
\tilde\h = -\log_{M}\Big[ 1 - v_{eff}\log M
\frac{2e^{2}}{5\pi^2}\big( 1 +A''_{-\io} \big) \Big]\;;\label{tildeh}
\ee
at lowest order, Eq.(\ref{tildeh}) gives $\tilde\h^{(2)}=2e^2/(5\p^2)$.

Similarly $C_1 M^{\h h}\le Z_h\le C_2 M^{\h h}$ for two suitable positive constants 
$C_1,C_2$, with $\h=\lim_{h\to-\io}\log_M(1+
z_{0,h})$; at lowest order we find (see Appendix \ref{app2}):
\be \b^{z,(2)}_{h} = \frac{\log M}{4\p^2} \, (2
e^2_{1,h}-e^2_{0,h}v_h^{-2})
\frac{\x_h-\arctan\x_h}{\x_h^3}\;,\label{3a.19z}\\
\ee
so that $\h^{(2)}=\frac{e^{2}}{12\pi^2}$. 

Before we conclude this Section, let us 
briefly comment about the relation between $Z_{h}$, $v_{h}$ and the functions 
$Z(\kk)$ and $v(\kk)$ appearing in the main result, see (\ref{1.6}).
If $|\kk|=M^h$, we define $Z(\kk)=Z_h$ and $v(\kk)=v_h$; for general $|\kk|\le 1$, we let $Z(\kk)$
and $v(\kk)$ be smooth interpolations of these sequences. Of course, we can choose 
these interpolations in such a way that, if
$M^{h}\leq |\kk|\leq M^{h+1}$, 
\bea
&&\Big| \frac{Z(\kk)}{Z_{h}} - 1\Big| \le \Big|\frac{Z_{h+1}}{Z_{h}} - 1\Big|=O(
\eta \log M)\;,\nn\\
&&\Big| \frac{v(\kk) - v_{h}}{v_{eff}-v_h}\Big|\le 
\Big| \frac{v_{h+1} - v_{h}}{v_{eff}-v_h}\Big|=
O(\tilde\eta \log M)\;.\label{cont1}
\eea
therefore, we can replace in the leading part of the 2-point Schwinger function (\ref{4.12}) 
the wave function renormalization $Z_{j}$ and the effective Fermi velocity $v_j$ by
 $Z(\kk)$ and $v(\kk)$, provided that the correction term $\tilde B(\kk)$ in (\ref{4.12})
is replaced by a quantity $B(\kk)$ defined so to take into account higher order corrections 
satisfying the bounds (\ref{cont1}). This leads to the main result Eq.(\ref{1.6}).

\subsection{The flow of the effective charges at lowest order}

The physical behavior of the system is driven by the flow of
$e_{\m,h}$; in the following section, by using a WI relating the three- and two-point 
functions, we will show that $e_{\m,h}$ remain close to their 
initial values for all scales $h\le 1$ and $\lim_{h\to-\io}e_{\m,h}=e_{\m,-\io}=e+F_\m$,
where $F_\m$ can be expressed as series in the renormalized couplings 
starting at third order in the effective charges. In perturbation theory, 
this fact follows from non-trivial cancellations that are present at all orders. 
For illustrative purposes, here we perform
the lowest order computation in non-renormalized perturbation
theory, {\it in the presence of an infrared cutoff on the bosonic
propagator} $\c_{[h,0]}(\pp)=\c_0(\pp)-\c_0(M^{-h}\pp)$; at this lowest order, such a ``naive'' computation
gives the same result as the renormalized one; for the full
computation, see next section and Appendix \ref{app4}. If
\be
(\bar\g_0,\bar\g_1,\bar\g_2):=(\g_0,v\g_1,v\g_2)\;,\quad (\bar
k_0,\bar k_1,\bar k_2):=(k_0,v k_1,v k_2)\;,\label{gbar}
\ee
the effective charges on scale $h$ at third order are given by:
\be(e^{(3)}_{\m,h}-e)\g_\m = ie^{3}\int
\frac{d\kk}{(2\pi)^3}\, \frac{\c_{[h,0]}(\kk)}{2|\kk|}
\Big[\bar\g_\n g^{(\le
0)}(\kk)i\g_\m g^{(\le 0)}(\kk)\bar\g_\n + \bar\g_\n\partial_{\bar
k_\m}g^{(\le 0)}(\kk) \bar\g_\n\Big] \label{e3}\ee
where the first term in square brackets is the
vertex renormalization, while the second term is due to the wave
function and velocity renormalizations. Note that both integrals
are well defined in the ultraviolet (thanks to the presence of an
ultraviolet cutoff in the propagators), while for $h\to-\io$ they
are logarithmically divergent in the infrared. However, a
remarkable cancellation takes place between the two integrals; in
fact:
\bea &&g^{(\le 0)}(\kk)i\g_\m g^{(\le 0)}(\kk) + \partial_{\bar
k_\m} g^{(\le 0)} (\kk) =\label{5.14}\\&&\hskip3cm=
\frac{\partial_{\bar k_\m}\chi_0(\kk)}{i{\sl \kk}} +
\chi_0(\kk)\big(\chi_0(\kk) - 1\big)\frac{1}{i{\sl \kk}}i\g_\m
\frac{1}{i{\sl \kk}}\;,\nn\eea
with ${\sl \kk}:=k_0\g_0+v\vec k\cdot \vec \g$, so that
\bea &&e_{0,h}^{(3)} = e  + ie^{3}\int \frac{d\kk}{(2\pi)^3}\,
\bar\g_\m \frac{1}{i{\sl \kk}}
\bar\g_\m \frac{k_0}{2 |\kk|^2}\chi_0'(\kk)\chi_{[h,0]}(\kk) + O\big(e^3(M-1)\big) \;\label{5.16}\\
&&e_{1,h}^{(3)} = e + \frac{ie^3}{v} \int \frac{d\kk}{(2\pi)^3}\,
\bar\g_\m \frac{1}{i{\sl \kk}}\bar\g_{\m}
\frac{k_1}{2|\kk|^2}\chi_0'(\kk)\chi_{[h,0]}(\kk) + O\big(e^3(M-1)\big) \;.\nn
\eea
Notice that the cancellation {\it does not} depend on the presence 
of the bosonic IR cutoff; this fact will play an important role in the analysis 
at all orders of the flow of the effective charges, see next Section. 
An explicit computation of (\ref{5.16}) says that, at third order in $e$,
\be e_{\m,-\io}^{(3)} = e + e \a^{(2)}_\m\;,\label{4.9a}\ee
where
\bea &&\a_0^{(2)} = \frac{e^{2}}{8\pi^2}\big(2 - v^{-2}\big)\Big(
\frac{\x_0 - \arctan\x_0}{\x_0^{3}}\Big)+O(e^2(M-1))\;,\qquad
\label{5.6b}\\
&&\a_1^{(2)} = \frac{e^2}{16\pi^2}\frac{1}{v^2}\Big(
\frac{\arctan\x_0}{\x_0} - \frac{\x_0 - \arctan\x_0}{\x_0^3}
\Big)+O(e^2(M-1))\;;\qquad\label{5.7} \eea
the correction terms $O(e^2(M-1))$ can be made as small as
desired, by choosing $0<M-1\ll1$. Note that the two effective
charges are different:
\be e_{0,-\io}^{(3)}-e_{1,-\io}^{(3)}=-\frac{e^3}{5\p^2}F(v)+O\big(e^3(M-1)\big)
\;,\label{4.101a}\ee
where $F(v)$ the function defined in (\ref{1.88}). Combining (\ref{4.101a}) with 
(\ref{veff}) gives the last equation of (\ref{second})

Of course, the one-loop computation that we just described does not say much:
if we could not guarantee that a similar cancellation takes place at all orders
there would always be the possibility that higher orders produce a
completely different behavior, {\em e.g.} a vanishing or diverging
flow for $e_{\m,h}$, corresponding to completely different
physical properties of the system. In order to obtain a control at
all orders on $e_{\m,h}$ one needs to combine
the multiscale evaluation of the effective potentials with Ward
Identities. This is not a trivial task: Wilsonian RG methods are
based on a multiscale momentum decomposition which breaks the
local gauge invariance, which Ward identities are based on. In
Section \ref{sec5} below, following a strategy recently proposed
and developed in \cite{BM}, we
will prove (\ref{4.9a}).\\

{\bf Remark.} Note the unusual feature that $e_{0,h}\not=
e_{1,h}$, an effect due to the presence of the momentum cut-off
and the fact that $v\not=1$. The discussion of this and previous
sections can be repeated in the case that the bare interaction
involves two different charges, $e_0$ and $e_1$, describing the
couplings of the photon field with the temporal and spatial
components of the current. If $e=(e_0+e_1)/2$ and
$e_0-e_1=O(e^3)$, the conclusion is that $v_{eff}=1-(e^2/6\pi^2)
F(v)+(5/6)(e_0-e_1)/e+O(e^4)$ and it is of course possible to fine
tune the bare parameters $e_0$ and $e_1$ in such a way that
$e_{0,-\io}=e_{1,-\io}$ and $v_{eff}=1$. Note that, in a more
realistic model for graphene, describing tight binding electrons on the 
honeycomb lattice coupled with a 3D photon field via a lattice gauge 
invariant coupling, one expects that $e_{0,-\io}=e_{1,-\io}$ and $v_{eff}=1$.

\section{Ward Identities}\label{sec5}
\setcounter{equation}{0}
\renewcommand{\theequation}{\ref{sec5}.\arabic{equation}}

In this section we prove that order by order in perturbation
theory the effective charges $e_{\m,h}$ remain close to their
original values $e_{\m,0}=e$; moreover, we prove that asymptotically as
$h\to-\io$, $e_{0,h}\neq e_{1,h}$, see (\ref{4.9a})--(\ref{4.101a}).
The proof is based on a suitable
combination of the RG methods described in the previous sections
together with Ward Identities; even though the momentum
regularization breaks the
local gauge invariance needed to formally derive the WIs, we will
be able, following the strategy of \cite{BM}, to rigorously take
into account the effects of cutoffs, and to control the
corrections generated by their presence.

As anticipated at the end of Section \ref{sec3}, we consider a sequence of
models,
to be called {\it reference models} in what follows, with different infrared
bosonic cutoffs on scale $h$, {\em i.e.} with bosonic propagator given by:
\be
w^{[h,0]}(\pp) \equiv \frac{\chi_{[h,0]}(\pp)}{2|\pp|}\;,\quad
\chi_{[h,0]}(\pp) \equiv \chi_0(\pp) - \chi_0(M^{-h}\pp)\label{5.01}\ee
(the idea of introducing an infrared cutoff only in the bosonic sector is
borrowed from Adler and Bardeen \cite{AB}, who used a similar regularization
scheme in order to understand anomalies in quantum field theory).
The generating functional $\WW_{[h,0]}(J,\phi)$ of the correlations of the
reference model can be evaluated following an iterative procedure similar to
the one described in Section \ref{sec3} (see Appendix \ref{app3} for details),
with the important difference that after the integration of the scale $h$ we
are left with a purely fermionic theory, which is {\em superrenormalizable}: in
fact, setting $m=0$ in the formula for the scaling dimension (see lines
following (\ref{4.11a}) and recall that for scales smaller than $h$ the
reference model has no bosonic lines) we recognize that the scaling dimension
of this fermionic theory is $3-2n$, which is always negative once that the
two-legged subdiagrams have been renormalized, see \cite{GM}.
Let us denote by $\{e^{[h]}_{\m,k}\}_{k\geq h}$
the effective couplings of the reference model; of course, if $k\geq h$
\be
e^{[h]}_{\m,k}=e_{\m,k}\;,
\ee
where $\{e_{\m,k}\}_{k\leq 0}$ are the running coupling constants
of the original model. On the other hand, as proven in
Appendix \ref{app3}, the vertex functions
$\media{j_{\m,\pp};\psi_{\kk+\pp}\lis\psi_{\kk}}_h$ of the reference model
with bosonic cutoff on scale $h$ computed at external momenta
$\kk$, $\kk+\pp$ such that $|\kk+\pp|,|\kk|\simeq M^{h}$ and $|\pp|\ll M^h$ are
proportional to the charges $e_{\m,h}^{[h]}=e_{\m,h}$, see (\ref{4.12a});
therefore, if we get informations on the vertex functions of the reference
models, we automatically infer informations on the effective couplings
of the original model.

Such informations are provided by Ward Identities;
by performing the change of variables $\psi_\xx\to
e^{i\a_\xx}\psi_\xx$, $\lis\psi_{\xx}\to e^{-i\a_{\xx}}\lis\psi_{\xx}$
in the generating functional $\WW_{[h,0]}(J,\phi)$ of the reference model and
using that the Jacobian of this transformation is equal to $1$, see
\cite{BM}, we get:
\be e^{\WW_{[h,0]}(J,\phi)}=\int P(d\psi)P_{[h,0]}(dA)e^{-\int d\xx\,
\lis\psi_\xx (e^{-i\a_\xx} D e^{i\a_\xx}-D)\psi_\xx + V(A,\psi)+B(J,\phi
e^{-i\a})}\;,\label{5.1}
\ee
where $P_{[h,0]}(dA)$ is the gaussian integration with propagator (\ref{5.01})
and, if ${\sl\kk}=\g_0 k_0 +v \vec \g\cdot \vec k$, the pseudo-differential
operator $D$ is defined by: $$(D\psi)_\xx=\int_{\c(\kk)>0}
\frac{d\kk}{(2\p)^3}\,
\frac{e^{i\kk\xx}}{\c_0(\kk)}i{\sl{\kk}}\,\psi_\kk\;.$$
If we derive (\ref{5.1})
with respect to $\a$, $\lis\phi$ and $\phi$ and then set $\a=\phi=J=0$,
we get the following identity:
\be p_{\m}\media{j_{\m,\pp};{\psi}_{\kk+\pp}\lis\psi_{\kk}}_h =
\media{\psi_{\kk}\lis\psi_{\kk}}_{h}
- \media{\psi_{\kk+\pp}\lis\psi_{\kk+\pp}}_h +\D_h(\kk,\pp) \label{5.2} \ee
where
\be \D_h(\kk,\pp)=\int \frac{d\kk'}{(2\p)^3}\,
\media{\lis\psi_{\kk' + \pp}C(\kk',\pp)\psi_{\kk'};\psi_{\kk+\pp}
\lis\psi_\kk}_h\label{5.3} \ee
and
\be C(\kk,\pp)=
i{\sl\kk}\Big(\chi_0(\kk)^{-1}-1\Big)-i({\sl\kk}+{\sl\pp})\Big(\chi_0(\kk+
\pp)^{-1}-1\Big)\;. \label{5.4}\ee
The correction term $\D_h(\kk,\pp)$ in (\ref{5.2}) is due to the
presence of the ultraviolet momentum cut-off, and it can be
computed by following a strategy analogous to the one used to
prove the vanishing of the beta function in one-dimensional Fermi
systems \cite{BM}. We can write
\be \D_h(\kk,\pp)= \a_{\m} p_\m
\media{j_{\m,\pp};\psi_{\kk+\pp}\lis\psi_{\kk}}_h+
\frac{p_\m}{Z_h} R_{\m,h}(\kk,\pp) \;,\label{5.6} \ee
where the correction $R_{\m,h}(\kk,\pp)$ is dimensionally negligible 
with respect to the first term, see Appendix \ref{app4}. More precisely, in Appendix \ref{app4}
it is shown that: (i) $R_{\m,h}(\kk,\pp)$ can be written as a sum over trees 
with $N$ endpoints of contributions $R_{\m,h}^{(N)}(\kk,\pp)$; (ii) it
is possible to choose $\a_\m$ in such a way that, under the same conditions
of Theorem \ref{thm1} and if $|\kk| = M^{h}$, $|\kk+\pp|\leq M^h$ and $|\pp|\ll M^h$,
\be |R_{\m,h}^{(N)}(\kk,\pp)|\le (\const.)^N\big(\frac{N}2\big)!
M^{-2h}M^{\frac{h}{2}}\bar\e_h^{N}\;.\label{restN}\ee
An explicit computation, see Appendix \ref{app4}, shows that at lowest order
$\a_{\m}^{(2)}$ is given by (\ref{5.6b})-(\ref{5.7}). 

Let us now show how to use the previous relations in order to derive bounds on 
the effective charges. Let us pick
$|\kk|=M^h$ and $|\pp|\ll M^h$; by using
Eqs.(\ref{4.12})-(\ref{4.12a}) and the fact that
\be g^{(h)}(\kk)-g^{(h)}(\kk+\pp) = g^{(h)}(\kk+\pp)(ip_0\g_0+iv_{h-1}
\vec p\cdot\vec \g)g^{(h)}(\kk)+p_{\m}\hat r_{\m}(\kk,\pp)\;,\label{5.7aa}\ee
with $\hat r_{\m}(\kk,\pp) = O(|\pp| M^{-3h})$, we find that
\bea
&&\media{\psi_{\kk}\lis\psi_{\kk}}_h - \media{\psi_{\kk+\pp}
\lis\psi_{\kk+\pp}}_h =  \frac1{Z_{h-1}} g^{(h)}(\kk+\pp)(ip_0\g_0+iv_{h-1}
\vec p\cdot\vec \g) g^{(h)}(\kk) +\nn\\&&\hskip4.2cm +\frac{p_{\m}}{Z_{h-1}}
\Big(\tilde r_{\m}(\kk,\pp) + \hat r_{\m}(\kk,\pp)\Big)\;,\label{5.8}\\
&&p_\m\media{j_{\m,\pp};\psi_{\kk+\pp}\lis\psi_{\kk}}_h = \frac{1}{eZ_{h-1}}
g^{(h)}(\kk+\pp)(ie_{0,h}p_0\g_0+iv_{h-1}e_{1,h}\vec p\cdot\vec \g)
g^{(h)}(\kk)+\nn\\&&\hskip3.3cm+\frac{p_\m}{Z_{h-1}}r_{\m}(\kk,\pp)\;,
\label{5.8b}\eea
with $|r_{\m}(\kk,\pp)|,|\tilde r_{\m}(\kk,\pp)|$ expressed by sums over trees of order 
$N\geq 2$ of contributions $r_{\m}^{(N)}(\kk,\pp), \tilde r_{\m}^{(N)}(\kk,\pp)$ 
bounded by (see Appendix \ref{app3}, formulas (\ref{A3.220}) and (\ref{A3.2302}))
\be|r_{\m}^{(N)}(\kk,\pp)|+|\tilde r_{\m}^{(N)}(\kk,\pp)|\le (\const.)^{N}\bar\e_{h}^{N} 
\big(\frac{N}{2}\big)!\, M^{-2h}\;.\label{estimate}\ee 
Now, if we plug (\ref{5.6}) into the Ward identity (\ref{5.2}), and we use the relations
(\ref{5.8})-(\ref{5.8b}), we get an identity that, computed at
$\kk=\kk_0 := (M^{h},\vec 0)$ and $\pp=\pp_0 := (p,\vec 0)$, after
taking the limit $p\to0$, reduces to:
\bea &&\frac{e_{0,h}}{e}(1 - \a_0) = 1 +
i M^{2h}\big[ \tilde r_{0}(\kk_0,\V0) + R_{0,h}(\kk_0,\V0) - (1 - \a_0)
r_{0}(\kk_0,\V0) \big]\g_0\nn\\&&\hskip2cm\equiv 1+A_{0,h}\;,\label{5.8c}\eea
with $A_{0,h}$ a sum of contributions associated to trees of order $N\geq 2$
bounded at the $N$-th order by $(\const.)^{N}(\bar\e_{h})^{N} (N/2)!$, 
as it follows from the estimates on $R_{0,h},r_0,\tilde r_0$ (note the crucial 
point that such estimate is proportional to $(\bar\e_{h})^N$ rather than to 
$(\bar\e_{-\io})^N$; this is the main reason why we chose to introduce the 
infrared cutoff on the bosonic propagator, see 
the end of Section \ref{schwing} and the beginning of this section). 
Eq.(\ref{5.8c}) combined with (\ref{5.6b})
implies, as desired, that {\it the effective charge $e_{0,h}$
remains close to $e_{0,0}=e$ at all orders in renormalized
perturbation theory}. Moreover, proceeding as in the derivation of 
Eq.(\ref{dec3}), we find that $|A_{0,h}-A_{0,-\io}|=O(e^2(v_{eff}-v_h))+O(e^2M^{\th h})$,
for some $0<\th<1$, from which we get Eq.(\ref{ch1}) for $\m=0$.
Similarly, if $\kk_1 := (0,M^{h},0)$, we
get:
\bea &&\frac{e_{1,h}}{e}(1 - \a_1) = \frac{v_{h-1}}{v_h} +
i M^{2h}v_{h-1}\cdot\nn\\&&\hskip3cm\cdot
\big[ \tilde r_{1}(\kk_1,\V0) + R_{1,h}(\kk_1,\V0) - (1 - \a_1)
r_{1}(\kk_1,\V0) \big]\g_1\nn\\&&\hskip2cm\equiv 1+A_{1,h}\;,\label{5.8d}\eea
with $A_{1,h}$ a sum of contributions associated to trees of order $N\geq 2$
bounded  at the $N$-th order 
by $(\const.)^{N}(\bar\e_{h})^{N} (N/2)!$, which implies that {\it
the effective charge $e_{1,h}$ remains close to $e_{1,0}=e$ at all
orders in renormalized perturbation theory}. Moreover, as in the $\m=0$ case, 
$|A_{1,h}-A_{1,-\io}|=O(e^2(v_{eff}-v_h))+O(e^2M^{\th h})$,
for some $0<\th<1$, from which we get Eq.(\ref{ch1}) for $\m=1$.

Equations (\ref{5.8c}) and (\ref{5.8d}) not only imply the
boundedness of the effective charges $e_{\m,h}$, but they also
allow us to compute the difference $e_{0,h}-e_{1,h}$, asymptotically as $h\to-\io$,
at all orders in renormalized perturbation theory. 
At lowest order, $e_{0,h}^{(3)} - e_{1,h}^{(3)} = e(\a_{0}^{(2)} -
\a_{1}^{(2)})$, as anticipated in previous section. 

\section{Conclusions}\label{sec6}
\setcounter{equation}{0}
\renewcommand{\theequation}{\ref{sec6}.\arabic{equation}}

We considered an effective continuum model for the low energy
physics of single-layer graphene, first introduced by Gonzalez et al. in
\cite{V1}. We analyzed it by {\it constructive Renormalization Group} methods,
which have already been proved effective in the {\it non perturbative} study of
several low-dimensional fermionic models, such
as one-dimensional interacting fermions \cite{BM}, or the
Hubbard model on the honeycomb lattice \cite{GM}. While in the present case
we are not able yet to prove the convergence of the renormalized
expansion, we can prove that it is {\it order by order finite}, see
Theorem \ref{thm1} above. Note that, on the contrary, the power
series expansion in the bare couplings is plagued by {\it
logarithmic divergences} and, therefore, informations obtained from it by
lowest order truncation are quite unreliable. In perspective, the proof
of convergence of the renormalized expansion appears to be much more
difficult than the one in \cite{BM} or \cite{GM}, due to the
simultaneous presence of bosons and fermions, but it should be
feasible (by using determinant bounds for the fermionic
sector and cluster expansion techniques for the boson sector).

A key point of our analysis is the control at all orders of the
flow of the effective couplings: this is obtained via Ward Identities relating
three- and two-point functions, by using a
technique developed in \cite{BM} for the analysis of Luttinger
liquids, in cases where bosonization cannot be applied (like in the
presence of an underlying lattice or of non-linear bands).
The Ward Identities have {\it corrections} with respect to the formal ones,
due to the presence of a fermionic ultraviolet cut-off. Remarkably, these
corrections can be rigorously bounded at all orders in renormalized
perturbation theory (see Section \ref{sec5}).

Several questions remain to be understood. First of all, the
effective model we considered is clearly not fundamental: a more
realistic model for graphene should be obtained by considering
electrons on the honeycomb lattice coupled to an electromagnetic
field living in the 3D continuum.
We believe that a Renormalization Group analysis, similar to the
one we performed here, is possible also for the lattice model, by
combining the techniques and results of \cite{GM} with those of
the present paper; we expect that the lattice model is asymptotic
to the continuum one considered here, provided that the bare
parameters of the continuum model are properly tuned. Another
important open problem is to understand the behavior of the system
in the case of static Coulomb interactions; this case can
be obtained by taking the limit $c\to\io$ together with a proper
rescaling of the electronic charge in the model with retarded
interactions. However, as discussed in the Remark at the end of
Section \ref{bound}, the static case seems to be much more
subtle than the one considered in this paper, since it apparently 
requires cancellations even to prove renormalizability of the 
theory at all orders. We plan to come
back to this case in a future publication.
\section*{Acknowledgements}
A.G. and V.M. gratefully acknowledge financial support from the
ERC Starting Grant CoMBoS-239694. We thank D. Haldane and
M. Vozmediano for many valuable discussions.

\appendix

\section{Symmetries}\label{appsim}
\setcounter{equation}{0}
\renewcommand{\theequation}{\ref{appsim}.\arabic{equation}}

In this Appendix we prove formulas (\ref{3.7}) and (\ref{3.8}); to do this, 
we exploit suitable symmetry transformations. We use the following 
explicit representation of the euclidean gamma matrices: 
\be
\g_0 = \begin{pmatrix} 0 & I \\ -I & 0 \end{pmatrix}\;,
\quad \g_1 = \begin{pmatrix} 0 & i\s_2 \\ i\s_2 & 0 \end{pmatrix}\;,
\quad \g_2 = \begin{pmatrix} 0 & i\s_1 \\ i\s_1 & 0 \end{pmatrix}\;,
\label{sim9b}
\ee
where 
$\s_1 = \begin{pmatrix} 0 & 1 \\ 1 & 0 \end{pmatrix}$ and 
$\s_2 = \begin{pmatrix} 0 & -i \\ i & 0 \end{pmatrix}$. It is also useful to 
define:  
$\g_3=\begin{pmatrix} 0 & -i\s_3 \\ -i\s_3 & 0 \end{pmatrix}$, with 
$\s_3 = \begin{pmatrix} 1 & 0 \\ 0 & -1 \end{pmatrix}$, and the corresponding 
fifth gamma matrix 
\be \g_5=\g_0\g_1\g_2\g_3= \begin{pmatrix} I & 0 \\ 0 & -I \end{pmatrix}\;,
\label{gamma5}\ee
which anticommutes with all the other gamma matrices: $\{\g_\m,\g_5\}=0$,
$\forall \m=0,\ldots,3$. Finally, given $\o\in\{+,-\}$, we define the chiral 
projector $P_\o:=(1+\o\g_5)/2$.

It is straightforward to check that both the gaussian integrations $P(d\psi)$, 
$P(dA)$ and the interaction $V(A,\psi)$ are invariant under the following 
symmetry transformations, which are preserved by the multiscale integration:
\begin{itemize}
\item[(1)] \underline{Chirality:} 
$P_\o\psi_\kk\rightarrow e^{-i\a_\o}P_\o\psi_\kk$, 
$\lis\psi_\kk P_{-\o}\rightarrow \lis\psi_\kk P_{-\o}e^{+i\a_\o}$, 
with $\a_{\o}\in \RRR$ independent of $\kk$.
\item[(2)] \underline{Spatial rotations:} 
$\psi_{\kk}\rightarrow 
e^{\frac{\theta}{4}[\g_1,\g_2]}\psi_{R^{[1,2]}_{-\theta}\kk}$, 
$\overline \psi_{\kk}\rightarrow \overline \psi_{R^{[1,2]}_{-\theta}\kk} 
e^{- \frac{\theta}{4}[\g_1,\g_2]}$ and $A_{\m,\pp}\rightarrow 
\big[R^{[1,2]}_{\th}A_{\cdot,R^{[1,2]}_{-\th}\pp}\big]_{\m}$, with
\be
R^{[1,2]}_{\theta} = \begin{pmatrix} 1&0 &0\\ 
0&\cos\theta & -\sin\theta  \\ 
0& \sin\theta & \cos\theta \\\end{pmatrix}\;.\lb{sim12}
\ee
The invariance of the model under (2) is a simple 
consequence of the fact that 
\be
e^{- \frac{\theta}{4}[\g_1,\g_2]}\left( \g_0,\g_1,\g_2 \right)
e^{\frac{\theta}{4}[\g_1,\g_2]} = 
(\g_0, \g_1\cos \theta - \g_2\sin\theta, \g_2\cos\theta +\g_1\sin\theta)\;.
\label{sim12b}\ee
\item[(3)] \underline{Complex conjugation:} $\psi_\kk\to (-i\g_2)\psi_{-\kk}$,
$\lis\psi_{\kk}\to\lis\psi_{-\kk}(i\g_2)$, 
$A_{\m,\kk}\rightarrow - A_{\m,-\kk}$ and
$\kappa \rightarrow \kappa^*$, where $\kappa$ is a generic constant 
appearing in $P(d\psi)$, $P(dA)$ and/or in $V(A,\psi)$.
\item[(4.a)] \underline{Horizontal reflections:} $\psi_\kk\to
(i\g_3\g_1)\psi_{\tilde\kk}$, $\lis\psi_\kk\to\lis\psi_{\tilde\kk}
(-i\g_1\g_3)$ and $A_{\m,\pp}\rightarrow (-1)^\m A_{\m,\tilde\pp}$, 
where $\tilde\kk = (k_0,-k_1,k_2)$.
\item[(4.b)] \underline{Vertical reflections:} $\psi_\kk\to 
(-i\g_2)\psi_{\tilde\kk}$, $\lis\psi_{\kk}\to\lis\psi_{\tilde\kk}
(i\g_2)$ and $A_{\m,\pp}\rightarrow (-1)^{\d_{\m,2}}A_{\m,\tilde\pp}$, 
where $\tilde\kk = (k_0,k_1,-k_2)$.
\item[(5)] \underline{Particle -- hole:} $\psi_\kk\to (-\g_0\g_2)
\lis\psi^{T}_{\tilde\kk}$, $\lis\psi_\kk\to\psi^T_{\tilde\kk}\g_2\g_0$,
and $A_{\m,\pp}\rightarrow (-1)^{1-\d_{\m,0}}A_{\m,\tilde\pp}$, 
where $\tilde\kk = (k_0, -\vec k)$.
\item[(6)] \underline{Inversion:} $\psi_\kk\to \g_0\g_3\psi_{\tilde\kk}$,
$\lis\psi_\kk\to\lis\psi_{\tilde\kk}\g_3\g_0$ and 
$A_{\m,\kk}\rightarrow (-1)^{\d_{\m,0}}A_{\m,\tilde\pp}$,
where $\tilde\kk = (-k_0,\vec k)$.
\end{itemize}
In addition to the previous symmetries, if $v=c=1$ the theory has an additional
space-time invariance, namely:
\begin{itemize}
\item[(7)] \underline{Relativistic invariance:} $\psi_{\kk}\rightarrow 
e^{\frac{\theta}{4}[\g_0,\g_1]}\psi_{R^{[0,1]}_{-\theta}\kk}$, 
$\overline \psi_{\kk}\rightarrow \overline \psi_{R^{[0,1]}_{-\theta}\kk} 
e^{- \frac{\theta}{4}[\g_0,\g_1]}$ and $A_{\m,\pp}\rightarrow 
\big[R_{\th}A_{\cdot,R_{\th}^{-1}\pp}\big]_{\m}$, with
\be
R^{[0,1]}_{\theta} = \begin{pmatrix} \cos\theta & -\sin\theta & 0 \\ 
\sin\theta & \cos\theta & 0 \\ 0 & 0 & 1\end{pmatrix}\;.\lb{sim2}
\ee
\end{itemize}
The invariance of the model under (7) is a simple 
consequence of the remark that
\be
e^{- \frac{\theta}{4}[\g_0,\g_1]}\left( \g_0,\g_1,\g_2 \right)
e^{\frac{\theta}{4}[\g_0,\g_1]} = 
(\g_0\cos \theta - \g_1\sin\theta, \g_1\cos\theta +\g_0\sin\theta, \g_2)\;.
\label{sim2b}
\ee
It is now straightforward to check that these symmetries imply 
(\ref{3.7}), (\ref{3.8}). In fact, 
the first two identities in the first line of 
(\ref{3.7}) and the second identity in the second line of (\ref{3.7})
easily follow from (4.a)+(4.b)+(6). 
Using (4.a)+(4.b)+(6) we also find that 
\be
W^{(h)}_{2,0,\m,\n}(\V0) = \d_{\m\n}W^{(h)}_{2,0,\m,\m}(\V0)\;,\label{sim3}
\ee 
while, from (2)+(3), we get
\be
W^{(h)}_{2,0,1,1}(\V0) = W^{(h)}_{2,0,2,2}(\V0)\;,\qquad 
W^{(h)}_{2,0,\m,\m}(\V0)\in \RRR\;,\label{sim4}
\ee
which imply the first identity in the second line of (\ref{3.7})
(notice that, if $v=c=1$, from (7) we also 
get that $W^{(h)}_{2,0,0,0}(\V0) = W^{(h)}_{2,0,1,1}(\V0)$).

Let us now consider the combination $\lis\psi_\kk W^{(h)}_{0,1}(\kk)\psi_\kk$. 
Using the fact that 
$\{I,\g_5,\{\g_j\}_{0\le j\le 3},\{\g_j\g_5\}_{0\le j\le 3},
\{\g_{j_1}\g_{j_2}\}_{0\le j_1<j_2\le 3} \}$ is a complete basis for the 
space of complex $4\times4$ matrices, we can rewrite it as:
\be \lis\psi_\kk W^{(h)}_{0,1}(\kk)\psi_\kk=
\lis\psi_\kk \Big\{c_0(\kk)I+\sum_{j=0}^3\big[c_1^{j}(\kk)\g_j+
c_{15}^{j}(\kk)\g_j\g_5\big]+\sum_{0\le j_1<j_2\le 5}c_2^{j_1j_2}(\kk)
\g_{j_1}\g_{j_2}+c_5(\kk)\g_5\Big\}\psi_\kk\;.\label{base}\ee
Now, using the invariance under (1), we find that, e.g.,
\be \lis\psi_\kk c_0(\kk)I\psi_\kk=\sum_{\o=\pm}\lis\psi_\kk P_\o c_0(\kk)IP_\o
\psi_\kk=e^{-2i\a_\o}\sum_{\o=\pm}\lis\psi_\kk P_\o c_0(\kk)IP_\o
\psi_\kk\;,\label{chir}\ee
for all $\a_\o\in\RRR$, which implies that $c_0(\kk)=0$; similarly, using the
invariance under (1) and the fact that $[\g_5,P_\o]=[\g_{j_1}\g_{j_2},P_\o]=0$,
$\forall 0\le j_1<j_2\le 3$, we find that 
$c_5(\kk)=0$
and $c_2^{j_1j_2} (\kk)=0$, $\forall 0\le j_1<j_2\le 3$. 
Therefore, 
\bea &&  \int \frac{d\kk}{(2\p)^3}\lis\psi_\kk W^{(h)}_{0,1}(\V0)\psi_\kk=
\int \sum_{j=0}^3\frac{d\kk}{(2\p)^3}
\lis\psi_\kk \big[c_1^{j}(\V0)\g_j+
c_{15}^{j}(\V0)\g_j\g_5\big]\psi_\kk\;,\label{base1}\\
 && \int\frac{d\kk}{(2\p)^3}\lis\psi_\kk\kk\dpr_\kk W^{(h)}_{0,1}(\V0)\psi_\kk=
\sum_{j=0}^3\int \frac{d\kk}{(2\p)^3}
\lis\psi_\kk \big[\kk\dpr_\kk c_1^{j}(\V0)\g_j+
\kk\dpr_\kk c_{15}^{j}(\V0)\g_j\g_5\big]\psi_\kk\;.\qquad
\label{base2}\eea
Let us first look at (\ref{base1}). Using the invariance under (4.a), 
we find that $\big[c_1^{j}(\V0)\g_j+
c_{15}^{j}(\V0)\g_j\g_5\big]=\g_3\g_1\big[c_1^{j}(\V0)\g_j+
c_{15}^{j}(\V0)\g_j\g_5\big]\g_1\g_3$, which implies that $c_1^1(\V0)=
c_1^3(\V0)=c_{15}^1(\V0)=
c_{15}^3(\V0)=0$. Using (2), we find that also $c_1^2(\V0)=c_{15}^2(\V0)=0$;
finally, using (6), we find that $c_1^0(\V0)=
c_{15}^0(\V0)=0$. This concludes the proof of the  
third identity in the first line 
of (\ref{3.7}).

Let us now look at (\ref{base2}). The terms proprtional to $k_0$ in the r.h.s.
of (\ref{base2}) are invariant under (2)+(4.a)+(4.b), which implies that 
$\dpr_{k_0}c_1^1(\V0)=\dpr_{k_0}c_1^2(\V0)=\dpr_{k_0}c_1^3(\V0)=
\dpr_{k_0}c_{15}^j(\V0)=0$. The terms proportional to $k_1$ are 
invariant (4.b)+(6), while the terms proportional to $k_2$ are 
invariant (4.a)+(6); combining these transformations with (2), we find that 
$\dpr_{k_1}c_1^0(\V0)=\dpr_{k_1}c_1^2(\V0)=\dpr_{k_1}c_1^3(\V0)=
\dpr_{k_1}c_{15}^j(\V0)=0$, that $\dpr_{k_2}c_1^0(\V0)=\dpr_{k_2}c_1^1(\V0)=
\dpr_{k_2}c_1^3(\V0)=
\dpr_{k_2}c_{15}^j(\V0)=0$, and that 
$\dpr_{k_1}c_1^1(\V0)=\dpr_{k_2}c_1^2(\V0)$. Therefore, 
\be  \int\frac{d\kk}{(2\p)^3}\lis\psi_\kk\kk\dpr_\kk W^{(h)}_{0,1}(\V0)
\psi_\kk=\int \frac{d\kk}{(2\p)^3}
\lis\psi_\kk \big[a_0k_0\g_0+a_1\vec k\cdot\vec \g\big]\psi_\kk\;,\qquad
\label{base3}\ee
for two suitable constants $a_0,a_1$. Using the invariance under (3), we
find that $a_0=iz_{0,h}$ and $a_1=iz_{1,h}$, with $z_{\m,h}\in\RRR$,
which concludes the proof of the first line of (\ref{3.8}) (of course, 
if $v=c=1$, then from (7) we also get that $z_{0,h} = z_{1,h}$, that is 
{\it the speed of light is not renormalized}). 

A completely analogous discussion can be repeated for the second line of 
(\ref{3.8}), but we will not belabor the details here.

\section{Multiscale integration for the correlation functions}\label{app3}
\setcounter{equation}{0}
\renewcommand{\theequation}{\ref{app3}.\arabic{equation}}

The multiscale integration used to compute the partition function
${\mathcal W}(0,0)$, described in Section \ref{sec3}, can
be suitably modified in order to compute the two and three-point correlation
functions in the reference model with bosonic infrared cutoff
on scale $h$, see (\ref{5.01}). We start by rewriting the two
and three point Schwinger functions in the following way:
\bea
\media{\psi_{\kk}\lis\psi_{\kk}}_{h^{*}} &=&
\frac{\partial^{2}}{\partial\lis\phi_{\kk}\partial \phi_{\kk}}
\WW_{[h^*,0]}(J,\phi)\big|_{J=\phi =0}\;,\label{A3.1}\\
\media{j_{\m,\pp};\psi_{\kk+\pp}\lis
\psi_\kk}_{h^{*}} &=& \frac{\partial^{3}}{\partial J_{\pp}
\partial\lis \phi_{\kk+\pp}\partial\phi_{\kk}}\WW_{[h^*,0]}(J,\phi)
\big|_{J=\phi =0}\nn\eea
where $\WW_{[h^*,0]}(J,\phi)$ is the generating function of the reference
model. The two point Schwinger function $\media{\psi_{\kk}\lis\psi_{\kk}}$
appearing in our main result is obtained as 
$\lim_{h^{*}\rightarrow-\infty} \media{\psi_{\kk}\lis\psi_{\kk}}_{h^{*}}$.

In order to compute $\WW_{[h^*,0]}(J,\phi)$, we proceed in a way analogous
to the one described in Section \ref{sec3}. We iteratively integrate the
fields $\psi^{(0)},A^{(0)}$, $\ldots$, $\psi^{(h+1)},A^{(h+1)}$, $\ldots$,
and after the integration of the first $|h|$ scales we are left with a
functional integral similar to (\ref{3.3}), but now involving new terms
depending on $J,\phi$. Let us first consider the case $h\geq h^{*}$; the
regime $h<h^{*}$ will be discussed later.
\vskip.1cm
{\it Case $h\geq h^{*}$.} We want to inductively prove that
\bea
&&e^{\WW_{[h^*,0]}(J,\phi)} = \label{A3.2}\\&&=e^{|\L|E_{h}+\SS^{(\ge h)}(J,\phi)}
\int P(d\psi^{(\le h)})P_{[h^*,0]}(dA^{(\le h)})
e^{\VV^{(h)}(A^{(\leq h)} + G_{A} J,\sqrt{Z_h} \psi^{(\leq h)})}
\cdot\nn\\
&&\hskip2cm\cdot e^{\BBB_{\phi}^{(h)}(A^{(\leq h)} + G_{A} J,
\sqrt{Z_h}\psi^{(\leq h)},\phi)+W_{R}^{(h)}(A^{(\leq h)} + G_{A} J,\sqrt{Z_h}
\psi^{(\leq h)},\phi)}\;,\nn
\eea
where: $\SS^{(\ge h)}(J,\phi)$ is independent of $(A,\psi)$,
$W_{R}^{(h)}$ contains terms explicitly depending on $(A,\psi)$ and of order
$\geq 2$ in $\phi$, while $\BBB_{\phi}^{(h)}$ is given by:
\bea
&&\BBB^{(h)}_{\phi}(A,\sqrt{Z_h}\psi,\phi) =\int \frac{d\kk}{(2\pi)^3}\,
\Big[\lis\phi_{\kk}[Q^{(h+1)}(\kk)]^{\dag}\psi_{\kk}  + \lis\psi_{\kk}
 Q^{(h+1)}(\kk)\phi_{\kk}\Big] + \nn\\&&
\quad+\int \frac{d\kk}{(2\pi)^3}\, \Big[ \lis \phi_{\kk} [G^{(h+1)}_{\psi}(\kk)
]^{\dag}\frac{\partial}{\partial\lis\psi_{\kk}}\VV^{(h)}(A,\sqrt{Z_h}\psi) +
\nn\\&&\qquad+\frac{\partial}{\partial\psi_{\kk}}\VV^{(h)}(A,\sqrt{Z_h}\psi)
G^{(h+1)}_{\psi}(\kk)\phi_{\kk} \Big]\;.\label{A3.3}\eea
Moreover, the functions $G_{A}$, $Q^{(h)}$, $G_{\psi}^{(h)}$ are defined
by the following relations:
\bea
&&eG_{A,\m}(\pp) := 1 + \n_{\m}w^{[h^{*},0]}(\pp)\;,\quad
G_{\psi}^{(h)}(\kk):= \sum_{i=h}^{0}\frac{g^{(i)}(\kk)}{Z_{i-1}}Q^{(i)}(\kk)
\;,\nn\\
&& Q^{(h)}(\kk):= Q^{(h+1)}(\kk) - iZ_{h}z_{\m,h}k_{\m}
\g_{\m}G_{\psi}^{(h+1)}(\kk)\;,\label{A3.4}
\eea
with $Q^{(1)}(\kk) \equiv 1$, $G^{(1)}(\kk) \equiv 0$. Note that, if $\kk$ is
in the support of $g^{(h)}(\kk)$,
\bea
&&Q^{(h)}(\kk) = 1 - iz_{\m,h}k_\m \g_\m g^{(h+1)}(\kk)\;,\nn\\
&&G_{\psi}^{(h)}(\kk) = \frac{g^{(h)}(\kk)}{Z_{h-1}}Q^{(h)}(\kk) +
\frac{g^{(h+1)}(\kk)}{Z_{h}}\;,\label{A3.3b}\eea
that is $||Q^{(h)}(\kk) - 1||\leq (\const.)\,\bar\e_h^{2}$ and
$||G_{\psi}^{(h)}(\kk)||\leq (\const.)\,Z_{h}^{-1}M^{-h}$.
Moreover, by the compact
support properties of $w^{[h^{*},0]}(\pp)$, $G_{A,\m}(\pp) \equiv e^{-1}$ for
all $|\pp|\leq M^{h^{*}}$.

In order to prove (\ref{A3.2})--(\ref{A3.4}) by induction, let us first
check them at the first step. The generating functional of the
correlations is defined as (see (\ref{1.1})-(\ref{1.2}))
\bea
&&e^{\WW_{[h^*,0]}(J,\phi)} =\label{A3.55}\\&&=\int P(d\psi^{(\leq 0)})P_{[h^*,0]}(dA^{(\leq 0)})
e^{\int \frac{d\pp}{(2\pi)^3}\,
(eA^{(\leq 0)}_{\m,\pp} + J_{\m,\pp})j_{\m,-\pp}^{(\leq 0)} -
\n_{\m}(A^{(\leq 0)}_\m,\,A^{(\leq 0)}_\m)+B(0,\phi)}\nn\eea
which, under the change of variables
\be
A^{(\leq 0)}_{\m,\pp}\rightarrow A^{(\leq 0)}_{\m,\pp} + e^{-1}\n_{\m}
w^{[h^{*},0]}(\pp) J_{\m,\pp}\;;\label{A3.5b}\ee
can be rewritten in the form (\ref{A3.2}), with
\bea
&&E_h=0\;,\quad e^2\SS^{(\ge 0)}=
\n_{\m}( J_\m,\, J_\m) + \n_{\m}^2(J_\m,\,w^{[h^{*},0]}J_\m)\;,\nn\\&&W_R^{(0)}=0\;,\quad
\VV^{(0)}=V\;,\quad \BBB_\phi^{(0)}(A + G_A J,\psi,\phi)=B(0,\phi)\;.
\eea
Let us now assume that (\ref{A3.2})--(\ref{A3.4}) are valid at scales
$\ge h$, and let us prove that the inductive assumption is reproduced at
scale $h-1$. We proceed as in Section \ref{sec3}; first, we renormalize
the free measure by reabsorbing into $\widetilde P(d\psi^{(\le h)})$ the
term $\exp\{\LL_\psi\VV^{(h)}\}$, see (\ref{3.9})--(\ref{3.12}), and then we
rescale the fields as in (\ref{3.13}). Similarly, in the definition
of $\BBB^{(h)}_{\phi}$, Eq.(\ref{A3.3}), we rewrite $\VV^{(h)}=
\LL_\psi\VV^{(h)}+\hat\VV^{(h)}$, combine the terms proportional to
$\LL_\psi\VV^{(h)}$ with those proportional to $Q^{(h+1)}$,
and rewrite
\bea&& \BBB^{(h)}(A,\sqrt{Z_h}\psi,\phi)=\hat \BBB^{(h)}(A,\sqrt{Z_{h-1}}\psi,
\phi):= \nn\\&&\quad:=\int \frac{d\kk}{(2\pi)^3}\,
\Big[\lis\phi_{\kk}[Q^{(h)}(\kk)]^{\dag}\psi_{\kk}  + \lis\psi_{\kk}
 Q^{(h)}(\kk)\phi_{\kk}\Big] + \nn\\&&
\qquad+\int \frac{d\kk}{(2\pi)^3}\, \Big[ \lis \phi_{\kk} [G^{(h+1)}_{\psi}(\kk)
]^{\dag}\frac{\partial}{\partial\lis\psi_{\kk}}\hat\VV^{(h)}(A,\sqrt{Z_{h-1}}
\psi) +\nn\\&&\quad\qquad+\frac{\partial}{\partial\psi_{\kk}}\hat\VV^{(h)}(A,\sqrt{Z_{h-1}}\psi)
G^{(h+1)}_{\psi}(\kk)\phi_{\kk} \Big]\;,\nn\eea
with $Q^{(h)}$ defined by (\ref{A3.4}). Finally, we rescale $W_{R}^{(h)}$,
by defining $$\hat W_{R}^{(h)}(A + G_A J,\sqrt{Z_{h-1}}\psi):=
W_{R}^{(h)}(A + G_A J,\sqrt{Z_h}\psi)\;,$$ and perform the integration
on scale $h$:
\bea
&&\int P(d\psi^{(h)})P(dA^{(h)})e^{\hat \VV^{(h)}(A^{(\leq h)} + G_A J,
\sqrt{Z_{h-1}}\psi^{(\leq h)})}\cdot\nn\\&&\hskip3cm\cdot e^{\hat\BBB_{\phi}^{(h)}(A^{(\leq h)} +
G_A J,\sqrt{Z_{h-1}}\psi^{(\leq h)}) + \hat W_{R}^{(h)}} \equiv \nn\\
&&\equiv e^{|\L|\tilde E_h + \SS^{(h-1)}(J,\phi)+
\VV^{(h-1)}(A^{(\leq h-1)} + G_A J,
\sqrt{Z_{h-1}}\psi^{(\leq h-1)})}\cdot\nn\\&&\hskip3cm\cdot e^{\BBB_{\phi}^{(h-1)}(A^{(\leq h-1)} +
G_A J,\sqrt{Z_{h-1}}\psi^{(\leq h-1)}) + W_{R}^{(h-1)}}\;,\nn
\eea
where $\SS^{(h-1)}(J,\phi)$ contains terms depending on $(J,\phi)$
but independent of $A^{(\le h-1)}$, $\psi^{(\le h-1)}$. Defining
$\SS^{(\ge h-1)}:=\SS^{(h-1)}+\SS^{(\ge h)}$, we immediately see that
the inductive assumption is reproduced on scale $h-1$.

{\it Case $h<h^{*}$}. For scales smaller than $h^{*}$,
there are no more bosonic fields to be integrated out, and we are left with a
purely fermionic theory, with scaling dimensions $3-2n$, $2n$ being the
number of external fermionic legs, see Theorem \ref{thm1} and following lines.
Therefore, once that the two-legged
subdiagrams have been renormalized and step by step reabsorbed into the free
fermionic measure, we are left with a superrenormalizable theory, as in
\cite{GM}. In particular, the four fermions interaction is irrelevant,
while the wave function renormalization and the Fermi velocity are modified
by a finite amount with respect to their values at $h^{*}$; that is, if
$\bar\e_{h^{*}} = \max_{k\geq h^{*}}\{|e_{\m,k}|,|\n_{\m,k}|\}$:
\be
Z_{h} = Z_{h^{*}}(1 + O(\bar\e_{h^{*}}^{2}))\;,\qquad v_{h} = v_{h^{*}}
(1 + O(\bar\e_{h^{*}}^{2}))\;.\label{A3.17b}
\ee
\vskip.2cm
{\it Tree expansion for the $2$-point function.}
As for the partition function (see section \ref{sec3}), the kernels of the
effective potentials produced by the multiscale
integration of $\WW_{[h^*,0]}(J,\phi)$ can be represented as sums
over trees, which in turn can be evaluated as sums over Feynman graphs.
Let us consider first the expansion for the $2$-point Schwinger
function. After having taken functional derivatives with respect
to $\phi_{\kk}$, $\lis\phi_{\kk}$ and after having set $J=\phi=0$,
we get an expansion in terms of a new class of trees
$\t\in\TT^{(h^*)}_{\bar k,\bar h, N}$,
with $\bar k\in(-\io,-1]$ the scale of the root and $\bar h>\bar k$;
 these trees are similar to the ones described in
section \ref{sec3},
up to the following differences.\\
(1) There are $N+2$ end--points and two of them, called
$v_{1},v_{2}$, are special and, respectively, correspond to
$\big[Q^{(h_{v_1}-1)}(\kk)\big]^{\dag}\psi_{\kk}^{(\le h_{v_1}-1)}$
or to $\lis\psi_{\kk}^{(\le h_{v_2}-1)} Q^{(h_{v_2}-1)}(\kk)$.\\
(2) The first vertex whose cluster contains both $v_{1}$, $v_{2}$, denoted by
$\bar v$, is on scale $\bar h$. No $\RR$ operation is associated to the
vertices on the line joining $\bar v$ to the root.\\
(3) There are no lines external to the cluster corresponding
to the root.\\
(4) There are no bosonic lines external to clusters on scale $h<h^*$.
\vskip.2cm
In terms of the new trees, we can expand the $2$-point Schwinger function as:
\be
\media{\psi_{\kk}\lis\psi_{\kk}}_{h^{*}} = \sum_{j=h_\kk}^{h_\kk+1}
[Q^{(j)}_{\psi}(\kk)]^\dagger
\frac{g^{(j)}(\kk)}{Z_{j-1}}Q^{(j)}(\kk) + \sum_{N=2}^{\infty}
\sum_{\bar h = -\infty}^{0}\sum_{\bar k = -\io}^{\bar h - 1}
\sum_{\t\in \TT^{(h^*)}_{\bar k,\bar h,N}}\SS_2(\t;\kk)\;,\label{A3.21}
\ee
where $h_{\kk}< 0$ is the integer such that $M^{h_{\kk}}\leq |\kk|<
M^{h_{\kk} + 1}$, and $\SS_2(\t;\kk)$ is defined in a way similar to
$\VV^{(h)}(\t)$ in (\ref{4.4}), modulo the modifications described in
items (1)-(4) above. Using the bounds described immediately after
(\ref{A3.3b}), which are valid for $\kk$ belonging to the support of
$g^{(h)}(\kk)$, and proceeding as in Section \ref{bound}, we get bounds on
$\SS_2(\t;\kk)$, which are the analogues of Theorem \ref{thm1}:
\be
\sum_{\bar h = -\infty}^{0}\sum_{\bar k = -\io}^{\bar h - 1}
\sum_{\t\in \TT_{\bar k,\bar h,N}^{(h^*)}}||\SS_2(\t;\kk)||\leq
(\const.)^{N}\bar\e_{h^*}^{N}\big(\frac{N}2\big)!
\frac{M^{-h_{\kk}}}{Z_{h_{\kk}}}\;.\label{A3.22}\ee
Notice that the result (\ref{A3.21}) and the 
bound (\ref{A3.22}) are true for any $\kk$ such that 
$|\kk|\geq M^{h^{*}}$; 
being the bound (\ref{A3.22}) uniform in $h^{*}$, 
our result on the two point function (\ref{1.6}) and (\ref{4.12}) is obtained
by fixing $\kk$ and taking the limit $h^{*}\rightarrow-\infty$ 
in (\ref{A3.21}).

In order to understand (\ref{A3.22}), it is enough to notice that,
as far as dimensional bounds are concerned, the vertices $v_{1}$
and $v_{2}$ play the role of two $\n$ vertices with an external
line (the $\phi$ line) and an extra
$Z_{h_{\kk}}^{-1/2}M^{-h_{\kk}}$ factor each. Moreover, since the
vertices on the path ${\mathcal P}_{r,\bar v}$ connecting the root
with $\bar v$ are not associated with any $\RR$ operation, we need
to multiply the value of the tree $\t\in\TT^{(h^*)}_{\bar k,\bar
h,N}$ by $M^{(1/2)(\bar h - \bar k)}M^{(1/2)(\bar k - \bar h)}$,
and to exploit the factor $M^{(1/2)(\bar k - \bar h)}$ in order to
renormalize all the clusters in ${\mathcal P}_{r,\bar v}$. Therefore,
\bea &&\sum_{\bar h = -\infty}^{0}\sum_{\bar k = -\io}^{\bar h - 1}
\sum_{\t\in \TT_{\bar k,\bar h,N}^{(h^*)}}||\SS_2(\t;\kk)||\leq
(\const.)^N\,\big(\frac{N}2\big)!\,\cdot\nn\\&&
\hskip2cm\cdot\frac{\bar\e_{h^*}^{N}}{Z_{h_{\kk}}}
 \sum_{\bar h\leq h_{\kk}}\sum_{\bar k\leq \bar h}
 M^{\bar k}M^{\bar h - h_{\kk}}M^{(1/2)(\bar h - \bar k)}
 M^{-2h_{\kk}}\qquad\label{A3.23}
\eea
where: the factor $M^{\bar k}$ is due to the fact that
graphs associated to the trees $\t\in \TT^{(h^*)}_{\bar k, \bar h, N}$
have two external lines; the factor $M^{\bar h - h_{\kk}}$ is
given by the product of the two short memory factors associated to the
two paths connecting $\bar v$ with $v_1$ and $v_2$, respectively; the
``bad'' factor $M^{(1/2)(\bar h - \bar k)}$ is the price
to pay to renormalize the vertices in ${\mathcal P}_{r,\bar v}$;
the $Z_{h_{\kk}}^{-1}$ and the
last $M^{-2h_{\kk}}$ are due to the fact that
$v_{1}$, $v_{2}$ behave dimensionally as $\n$
vertices times an extra $Z_{h_{\kk}}^{-1/2}M^{-h_{\kk}}$
factor. Performing the summation over $\bar k$ and $\bar h$ in (\ref{A3.23}),
we get (\ref{A3.22}). Note also that, if $\kk$ is
on scale $h_{\kk}\simeq {h^*}$, then the derivatives of $||\SS_2(\t;\kk)||$
can be dimensionally bounded as
\be \sum_{\bar h = -\infty}^{0}\sum_{\bar k = -\io}^{\bar h - 1}
\sum_{\t\in \TT_{\bar k,\bar
h,N}^{(h^*)}}||\dpr_\kk^n\SS_2(\t;\kk)||\leq
(\const.)^{N}\bar\e_{h^*}^{N}\big(\frac{N}2\big)!
\frac{M^{-(1+n)h_{\kk}}}{Z_{h_{\kk}}}\;,\label{A3.220}\ee
from which the bound on $\tilde r_\m^{(N)}(\kk,\pp)$ stated in
(\ref{estimate}) immediately follows. 

\vskip.2cm {\it Tree expansion for the $3$-point function.} 
Let us pick $|\kk|=M^{h^*}$, $|\kk+\pp|\le M^{h^*}$ and $|\pp|\ll M^{h^*}$,
which is the condition that we need in order to apply Ward
Identities in the form described in Section \ref{sec5}. In this
case, the expansion of $3$-point function
$\media{j_{\m,\pp};\psi_{\kk+\pp}\lis\psi_\kk}_{h^{*}}$ is very
similar to the one just described for the $2$-point function. The
result can be written in the form
\bea &&\media{j_{\m,\pp};\psi_{\kk+\pp}\lis\psi_\kk}_{h^{*}} =i\frac{\bar
e_{\m,h^{*}}}{e}[G^{(h^{*}-1)}_{\psi}(\kk+\pp)]^\dagger\g_\m
g^{(h^{*})}(\kk)Q^{(h^*)}(\kk) + \nn\\&& \hskip3cm+ \sum_{\substack{N\ge 1,\\\bar h
\le h^*}}\sum_{\substack{\bar k<\bar h,\\ h_{v_3}>h^*}}
\sum_{\t\in \TT_{\bar k,\bar h,h_{v_3},
N}^{(h^*)}}\SS_3(\t;\kk,\pp)\label{A3.24} \eea
where $\TT_{\bar k,\bar h,h_{v_3},N}^{(h^*)}$ is a new class of trees,
with $\bar k<0$ the scale of the root, similar to the trees in
$\TT_{\bar k,\bar h,N}^{(h^*)}$, up to the fact that they have
$N+3$ endpoints rather than $N+2$ (see item (1) in the list preceding
(\ref{A3.21})); three of them are special: $v_1$ and $v_2$
are associated to the same contributions described in item (1) above, while
$v_3$ is associated to a contribution $Z_{h_{\bar v_3}-1}\,
(e_{\m,h_{\bar v_3}}/e)
j^{(\le h_{\bar v_3})}_{\m,\pp}-M^{h_{\bar v_3}}(\n_{\m,h_{\bar v_3}}/e)
A_{\m,\pp}$, with $\bar v_3$ the vertex immediately
preceding $v_3$ on $\t$ (which the endpoint $v_3$ is attached to) and
$h_{v_3}>h^*$. The value of the tree, $\SS_3(\t;\kk,\pp)$, is defined in a way
similar to $\SS_2(\t;\kk)$, modulo the modifications described above.
$\SS_3(\t;\kk,\pp)$ admits bounds analogous to (\ref{A3.22})-(\ref{A3.23});
recalling that $|\kk|=M^{h^*}$, $|\kk+\pp|\le M^{h^*}$ and $|\pp|\ll M^{h^*}$,
we find:
\bea&&\sum_{\bar h = -\infty}^{h^*}\sum_{\bar k = -\io}^{\bar h - 1}
\sum_{h_{v_3}=h^*+1}^1
\sum_{\t\in \TT_{\bar k,\bar h,h_{v_3},N}^{(h^*)}}||\SS_3(\t;\kk,\pp)||\leq
(\const.)^N\,\big(\frac{N}2\big)!\,\bar\e_{h^*}^{N}\frac{1}{Z_{h^*-1}}\cdot
\nn\\
&&\hskip2cm\cdot\sum_{\substack{\bar h\leq h^*\\\bar k< \bar h\\ h_{v_3}>h^*}}
M^{(1/2)(\bar k-\bar h)}M^{\bar h - h^*}M^{(1/2)(h^*-h_{v_3})}
M^{-2h^*}\;,\label{A3.2301}
\eea
where: $M^{(1/2)(\bar k-\bar h)}$ is the short memory factor
associated to the path between the root and $\bar v$; $M^{\bar h -
h^*}$ is the product of the two short memory factors associated to
the paths connecting $\bar v$ with $v_1$ and $v_2$, respectively;
$M^{(1/2)(h^*-h_{v_3})}$ is the short memory factor associated to
a path between $h^*$ and $v_3$; $M^{-2h^*}/Z_{h^*-1}$ is the
product of two factors $M^{-h_\kk}Z_{h_\kk-1}^{-1/2}$ associated
to the vertices $v_1$ and $v_2$ (see the discussion following
(\ref{A3.22}) and recall that in this case $h_\kk=h^*$). We remark
that in this case, contrary to the case of the 2-point function,
the fact that there is no $\RR$ operator acting on the vertices on
the path between the root and $\bar v$ does not create any
problem, since those vertices are automatically irrelevant (they
behave as vertices with at least 5 external lines, i.e., $J$,
$\phi$, $\bar \phi$ and at least two fermionic lines) and,
therefore, $\RR=1$ on them. Note also that the vertices of type
$J\phi\psi$, which have an $\RR$ operator acting on, can only be
on scale $h^*-1$ or $h^*$ (by conservation of momentum) and,
therefore, the action of the $\RR$ operator on such vertices
automatically gives the usual dimensional gain of the form
$\const.\,M^{h_{v}-h_{v'}}$. Performing the summations over $\bar
k,\bar h, h_{v_3}$ in (\ref{A3.2301}), we find the analogue of
(\ref{A3.22}):
\be \sum_{\bar h = -\infty}^{h^*}\sum_{\bar k = -\io}^{\bar h - 1}
\sum_{h_{v_3}=h^*+1}^1
\sum_{\t\in \TT_{\bar k,\bar h,h_{v_3},N}^{(h^*)}}||\SS_3(\t;\kk,\pp)||\leq
(\const.)^N\,\big(\frac{N}2\big)!\,\bar\e_{h^*}^{N}\frac{M^{-2h^*}}{Z_{h^*-1}}
\;,\label{A3.2302}\ee
from which the bound on $r_\m^{(N)}(\kk,\pp)$ stated in (\ref{estimate}).

\section{Lowest order computations}\label{app2} \setcounter{equation}{0}
\renewcommand{\theequation}{\ref{app2}.\arabic{equation}}

In this Appendix we reproduce the details of the second order computations
leading to (\ref{3a.19n}), (\ref{3a.19nn}), (\ref{3a.19z}), (\ref{3a.19v}).
\subsection{Computation of $\b_h^{z,(2)}$}

By definition, see (\ref{3.8}) and (\ref{3a.18}),
$\b_h^{z}=z_{0,h}=-i\g_0\dpr_{k_0}W^{(h)}_{0,1}(\V0)$. At
one-loop, defining $\bar e_{0,h}=e_{0,h}$ and $\bar e_{1,h}= v_{h-1}
e_{1,h}$, we find:
\bea &&\b_h^{z,(2)}=z_{0,h}^{(2)} = -i\g_0\bar e_{\m,h+1}^2
\int\frac{d\pp}{(2\p)^3}\dpr_{p_0}\Big(\frac{f_{h+1}(\pp)}{2|\pp|}\Big)\g_\m
 g^{(h+1)}(\pp)\g_\m + \label{A1.1}\\&& \hskip1cm - i\g_0\bar e_{\m,h+2}^2\Big(
\frac{Z_{h+1}}{Z_h}\Big)^2\int\frac{d\pp}{(2\p)^3}\dpr_{p_0}\Big(
\frac{f_{h+2}(\pp)}{2|\pp|}\Big)\g_\m g^{(h+1)}(\pp)\g_\m + \nn\\&& \hskip1cm -
i\g_0\bar e_{\m,h+2}^2\frac{Z_{h+1}}{Z_h}\int\frac{d\pp}{(2\p)^3}
\dpr_{p_0}\Big(\frac{f_{h+1}(\pp)}{2|\pp|}\Big)\g_\m
 g^{(h+2)}(\pp)\g_\m\;.\nn\eea
Using inductively the beta function equations for $Z_{h+1}, v_{h+1},
e_{\m,h+2}$, and neglecting higher order terms, we can rewrite (\ref{A1.1})
as
\bea &&z_{0,h}^{(2)}=i\g_0\bar e_{\m,h+1}^2\frac{1}2
\int\frac{d\pp}{(2\p)^3}\frac{p_0^2}{|\pp|^3}\frac{i\g_\m\g_0\g_\m}{p_0^2+
v_h^2|\vec p|^2}\cdot\nn\\&&\hskip1cm\cdot\Big[(f_{h+1}(\pp)-|\pp|f_{h+1}'(\pp))(f_{h+1}(\pp)+
f_{h+2}(\pp))+\nn\\
&&\quad\hskip1cm+(f_{h+2}(\pp)-|\pp|f_{h+2}'(\pp))f_{h+1}(\pp)\Big]\;.
\label{A1.2}\eea
Passing to radial coordinates, $\pp=p(\cos\th,\sin\th\cos\ph,\sin\th\sin\ph)$,
and using the fact that $\int dp (f_{h+1}' f_{h+1}+f_{h+1}' f_{h+2}+
f_{h+2}' f_{h+1})=0$, we find:
\bea &&z_{0,h}^{(2)}=(2v_h^2e_{1,h}^2-e_{0,h}^2)\frac1{8\p^2}
\Big[\int_0^\io \frac{dp}{p}(f_{h+1}^2+2f_{h+1}f_{h+2})\Big]\cdot\nn\\
&&\hskip1cm\cdot\Big[\int_{-1}^1 d\cos\th\, \frac{\cos^2\th}{\cos^2\th+v^2_h\sin^2\th}
\Big]\;.\label{A1.3}\eea
The integral over the radial coordinate $p$
can be computed by using  the definition (\ref{3.1}):
\bea  &&\int_0^\io \frac{dp}{p}(f_{h+1}^2+2f_{h+1}f_{h+2}) = \label{A1.4}\\&&
=\int_0^\io \frac{dp}{p}[2(\c(p)-\c(Mp))-(\c^2(p)-\c^2(Mp))]
=\lim_{\e\to 0}\int_\e^{M\e}\frac{dp}p=\log M\;.\nn\eea
Finally, an explicit evaluation of the integral over $d\cos\th$ leads to
(\ref{3a.19z}).

\subsection{Computation of $z_{1,h}^{(2)}$}

By definition, see formulas (\ref{3.8}) and (\ref{3a.18}),
$\b_h^{v,(2)}=z_{1,h}^{(2)}-v_hz_{0,h}^{(2)}$, with
$z_{1,h}=-i\g_1\dpr_{k_1}W^{(h)}_{0,1}(\V0)$. At second order,
proceeding as in the derivation of (\ref{A1.2}), we find:
\bea &&z_{1,h}^{(2)}=i\g_1\bar e_{\m,h+1}^2\frac{1}2
\cdot\label{A1.5}\\&&\quad\cdot\int\frac{d\pp}{(2\p)^3}\frac{p_1^2}{|\pp|^3}
\frac{i\g_\m v_h\g_1\g_\m}{p_0^2+
v_h^2|\vec p|^2}\Big[(f_{h+1}(\pp)-|\pp|f_{h+1}'(\pp))(f_{h+1}(\pp)+
f_{h+2}(\pp))+\nn\\
&&\qquad+(f_{h+2}(\pp)-|\pp|f_{h+2}'(\pp))f_{h+1}(\pp)\Big]=
\nn\\
&&=e_{0,h}^2v_h\frac1{16\p^2}
\Big[\int_0^\io \frac{dp}{p}(f_{h+1}^2+2f_{h+1}f_{h+2})\Big]\cdot\Big[\int_{-1}^1 d\cos\th\, \frac{\sin^2\th}{\cos^2\th+v^2_h\sin^2\th}
\Big]\;.\nn\eea
An explicit evaluation of the integral leads to
\be z_{1,h}^{(2)}=e_{0,h}^2v_h^{-1}\frac{\log M}{8\p^2}
\Big(\frac{\arctan\x_h}{\x_h}-\frac{\x_h-\arctan\x_h}{\x_h^3}\Big)\;,
\label{A1.6}\ee
which, combined with $\b_h^{v,(2)}=z_{1,h}^{(2)}-v_hz_{0,h}^{(2)}$,
leads to (\ref{3a.19z}).

\subsection{Computation of $\b_{\m,h}^{\n,(2)}$}

By definition, see (\ref{3.7}) and (\ref{3a.18}), $\b^{\n}_{\m,h}
= -M^{-h+1}W^{(h-1)}_{2,0,\m,\m}(\V0) - M\n_{\m,h}$. At second
order, we find:
\bea
&&\b^{\n,(2)}_{\m,h} = -M^{-h+1}\frac{\bar e_{\m,h}^{2}}{2}
\int \frac{d\pp}{(2\pi)^3}\,\Tr\Big(\g_\m g^{(h)}(\pp)\g_\m g^{(h)}(\pp)\Big)
+ \label{A1.7}\\
&& \qquad\qquad- M^{-h+1}\bar e_{\m,h+1}^{2}\frac{Z_h}{Z_{h-1}}
\int \frac{d\pp}{(2\pi)^3}\,
\Tr\Big(\g_\m g^{(h+1)}(\pp) \g_\m g^{(h)}(\pp)\Big)\;.\nn
\eea
Using inductively the beta function equations for $e_{\m,h},Z_{h-1},v_{h-1}$,
 and neglecting higher orders, we can rewrite (\ref{A1.7}) as
\bea
&&\b^{\n,(2)}_{0,h} = -2M^{-h+1} e^{2}_{0,h}\int \frac{d\pp}{(2\pi)^3}
\frac{f_h(\pp)^2 + 2f_h(\pp)f_{h+1}(\pp)}{(p_0^2 + v_h^2 |\vec p|^2)^2}
(-p_0^2 +
v_h^2 |\vec p|^2)\;,\nn\\
&&\b^{\n,(2)}_{1,h} = -2M^{-h+1}\bar e^{2}_{1,h}\int \frac{d\pp}{(2\pi)^3}
\frac{f_h(\pp)^2 + 2f_h(\pp)f_{h+1}(\pp)}{(p_0^2 + v_h^2 |\vec p|^2)^2}p_0^2\;,
\label{A1.8}
\eea
where we used that $\Tr\big(\g_\m \g_\a \g_\m \g_\a \big) = -4$ if $\m\neq \a$
and $4$ otherwise; passing to radial coordinates we find
\bea
&&\b^{\n,(2)}_{0,h} = \frac{-2}{(2\pi)^2}M^{-h+1}e_{0,h}^2
\Big[\int_{0}^{\infty}dp\, \big( f_h^2 + 2f_h f_{h+1} \big)\Big]\cdot\nn\\&&
\qquad\qquad\cdot\int_{-1}^{1} d\cos\th\, \frac{-\cos^2\th + v_h^2\sin^2\th}{(\cos^2\th +
v_h^2\sin^2\th)^2}\;,\nn\\
&&\b^{\n,(2)}_{1,h} = \frac{-2}{(2\pi)^2}M^{-h+1}\bar e_{1,h}^2
\Big[\int_{0}^{\infty}dp\,\big( f_h^2 + 2f_h f_{h+1} \big)\Big]\cdot\nn\\&&
\qquad\qquad\cdot\int_{-1}^{1}d\cos\th\, \frac{\cos^2\th}{(\cos^2\th + v_h^2\sin^2\th)^2}\;.
\label{A1.9}
\eea
The integral over the radial coordinate $p$ can be rewritten as, using
 the definition (\ref{3.1}):
\be
\int_{0}^{\infty} dp\, \big(f_h^2 + 2f_h f_{h+1}\big) = M^{h-1}(M-1)
\int_{0}^{\infty} dp\, \big(2\chi(p) - \chi^2(p)\big)\;.\label{A1.9b}
\ee
Finally, an explicit evaluation of the integral over $d\cos\th$ leads
to (\ref{3a.19n}).

\section{Multiscale integration of the correction term to the WI}\label{app4}
\setcounter{equation}{0}
\renewcommand{\theequation}{\ref{app4}.\arabic{equation}}

In this Appendix we prove (\ref{5.6}) and the bound (\ref{restN}).
We assume that $h=h^*$, $|\kk|=M^h$ and $|\pp|\ll M^h$.
We start by rewriting
\be \frac{p_{\m}}{Z_{h}}R_{\m,h^{*}}(\kk,\pp) = \frac{\partial^{3}}{\partial
\tilde J_{\pp}\partial\lis\phi_{\kk+\pp}\partial\phi_{\kk}}\widetilde
\WW_{[h,0]}
(\tilde J,\phi)\big|_{\tilde J=\phi = 0}\;, \label{A4.2}
\ee
with $\widetilde \WW_{[h,0]}(\tilde J,\phi)$ defined as:
\be e^{\widetilde \WW_{[h,0]}(\tilde J,\phi)}:= \int P(d\psi)P_{[h,0]}(dA)\,
e^{V(A,\psi) + \widetilde B(\tilde J,\phi)}\;, \label{A4.3} \ee
and
\bea \widetilde B(\tilde J,\phi) &=& \int \frac{d\pp}{(2\pi)^3}\, \tilde J_{\pp}
\Big[
\int \frac{d\kk}{(2\p)^3}\,\lis\psi_{\kk+\pp}C(\kk,\pp)
\psi_\kk - \a_\m p_{\m} j_{\m,\pp} \Big]  + \nn\\&& + \int
\frac{d\kk}{(2\pi)^3}\, \big[\phi_{\kk}\lis \psi_{\kk} +
\lis\phi_{\kk}\psi_{\kk}\big] \label{A4.4}\;. \eea
The main difference with respect to the generating functional of the
correlation functions is the presence of the correction term proportional to
$C(\kk,\pp)$, see (\ref{5.4}) for a definition. Eq.(\ref{A4.3}) can again
be studied by RG methods, see \cite{BM} for further details.
A crucial role is played by the properties of the
function $C(\kk,\pp)$; it is easy to verify that
\be  g^{(i)}(\kk+\pp)C(\kk,\pp) g^{(j)}(\kk)\label{5.22}
\ee
is non vanishing only if at least one of the indices $i,j$ is equal to $0$;
moreover, when it is nonvanishing, it is dimensionally bounded from above
by $(\const.)|\pp|M^{-i-j}$.

\begin{figure}[htbp]
\centering
\includegraphics[width=0.9\textwidth]{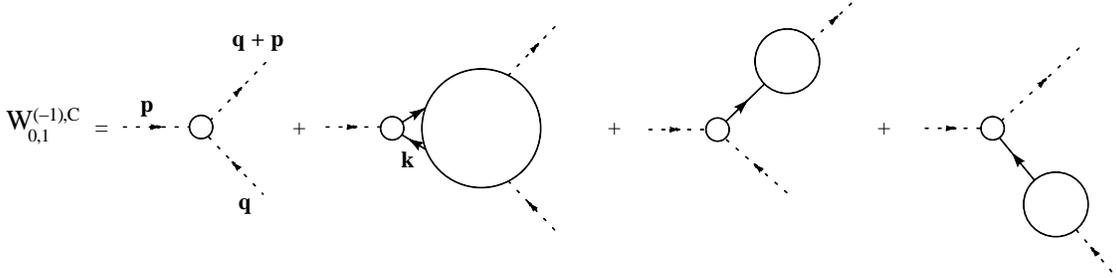}
\caption{Schematic representation of the expansion for
$W^{(-1),C}_{0,1}$; the small circle represents $C(\kk,\pp)$.} \label{fig3.1}
\end{figure}

We start by integrating the scale $0$, and we find:
\bea &&e^{\widetilde \WW_{[h,0]}(\tilde J,\phi)}= e^{|\L|E_{-1}+\widetilde
{\mathcal S}^{(\ge -1)}(\tilde J, \phi)}\cdot\label{A4.4b}\\&&\quad\cdot\int P(d\psi^{(\le -1)})P_{[h,-1]}(dA^{(\le -1)})\,e^{\VV^{(-1)}(A^{(\leq -1)},\sqrt{Z_{-1}}\psi^{(\leq -1)})
+\widetilde \BBB^{(-1)}}\;,\nn\eea
where $\widetilde{\mathcal S}^{(\ge -1)}$ collects the terms depending on
$\tilde J,\phi$ but independent of $A,\psi$, and
\be \widetilde \BBB^{(-1)}(A,\psi) = \widetilde \BBB_{J}^{(-1)}(A,\psi,\phi) +
\BBB^{(-1)}_{\phi}(A,\psi) + \widetilde W_{R}^{(-1)}\;,\label{A4.4c}
\ee
with: $\widetilde \BBB_{J}^{(-1)}(A,\psi,\phi)$ linear in $\tilde J$ and
independent of $\phi$; $\BBB^{(-1)}_{\phi}(A,\psi)$ given by (\ref{A3.3});
$\widetilde W_{R}^{(-1)}$ the rest, which is at least quadratic in
$(\tilde J,\phi)$. With respect to the computation of
$\WW_{[h,0]}(\tilde J,\phi)$, we now have new marginal terms of the form
$\tilde J \,\bar \psi\psi$, which are contained in $\widetilde
\BBB_{J}^{(-1)}(A,\psi,\phi)$ and need to be renormalized. Let us simbolically
represent by $\widetilde W_{m,n}^{(-1)}(\{\kk_i\},\{\qq_i\},\pp)$ the
generic non-trivial kernel appearing in $\widetilde \BBB_{J}^{(-1)}(A,\psi,
\phi)$; $m$ is the number of bosonic external lines (of either $\tilde J$ or
$A$ type) while $2n$ is the number of $\psi$ fields; $\{\kk_i\},\{\qq_i\}$ are
respectively the fermionic/bosonic momenta and
$\pp$ is the momentum flowing through $\tilde J$. As usual, these new
kernels can be represented as sums over Feynman graphs. The $\tilde J$ external
line can be attached to a simple vertex, corresponding to the monomial
$-\a_{\m}p_{\m}\tilde J_{\pp}j_{\m,\pp}^{(\leq 0)}$, or to a ``thick'' vertex,
representing $\tilde J_{\pp}\lis\psi_{\kk+\pp}C(\kk,\pp)\psi_{\kk}$ (the
``small circle'' associated to the vertex represents the matrix kernel
$C(\kk,\pp)$, see Fig.\ref{fig3.1}). Let us
denote by $W^{(-1),C}_{m,n}$ the contribution to $\widetilde W_{m,n}^{(-1)}$
coming from graphs with the $\tilde J$ line attached to a thick vertex, see
Figure \ref{fig3.1}.
By the properties of the $C(\kk,\pp)$ function, see \cite{BM} for
details, it follows that $\widetilde W_{m,n}^{(-1)}(\{\kk_i\},\{\qq_i\},\pp)
=: p_{\m}  \bar W_{m,n,\m}^{(-1)}(\{\kk_i\},\{\qq_i\},\pp)$,
with $\bar W_{m,n,\m}^{(-1)}$ dimensionally bounded as an
$A\lis\psi\psi$ kernel, uniformly in $\pp$. We define the
action of the $\RR\equiv 1 - \LL$ operator on $\bar W_{m,n,\m}^{(-1)}$ in a
way similar to (\ref{3.6})--(\ref{3.8}). In particular,
$\LL \bar W_{0,1,\m}^{(-1)}(\kk,\pp) := \bar W_{0,1,\m}^{(-1)}
(\V0,\V0)$ and, by symmetry,
\be Z_{-1}\tilde J_{\pp}\lis\psi_{\kk}\LL\bar W^{(-1)}_{0,1,\m}\psi_{\kk+\pp}
=- Z_{-2}\tilde J_{\pp} \a_{\m,-1}  j^{(\le -1)}_\pp\;,\label{A4.5}\ee
for a real constant $\a_{\m,-1}$, which is by definition the effective
$\a$-coupling on scale $-1$.
Note that the last two graphs in Fig.\ref{fig3.1} do not contribute to
$\a_{\m,-1}$ simply because they are one-particle reducible and, therefore,
they are vanishing at zero external momenta.

We now iterate the same procedure, and step by step the local parts of
the kernels of type $\tilde J\bar \psi\psi $ are collected together to form
a new running coupling constant, $\a_{\m,k}$; in order to show that
$R_{\m, h}$ is dimensionally negligible as $h\to-\io$, we need to show that
it is possible to fix the initial data $\a_\m=\a_{\m,0}$ in such a way that
$\a_{\m,h}$ goes exponentially to zero as $\h\to-\io$, which is proved in the
following.

{\it The flow of $\a_{\m,k}$.} The new marginal
running coupling constants $\a_{\m,h}$ evolve according to the flow equation:
$\a_{\m,k-1} = \a_{\m,k} + \b^{\a}_{\m,k}$, where
$\a_{\m,0} = \a_{\m}$ are the counterterms appearing in the bare
interaction (\ref{A4.4}). The beta function $\b^{\a}_{\m,h}$ can be split as
\be \b^{\a}_{\m,k} = \b^{\a,1}_{\m,k} + \b^{\a,2}_{\m,k}\;,\label{A4.9a}\ee
where $\b^{\a,1}_{\m,k}$ collects the contributions independent of $\a_{\m,k'}$
(which, therefore, are associated to graphs with the $\tilde J$ external line
emerging from the thick vertex representing
$C(\kk,\pp)$), and $\b^{\a,2}_{\m,k}$ collects the terms
from graphs with one vertex of type $\a_{\m,k'}$ for some $k'> k$.
It is crucial to recall that by the properties
of $C(\kk,\pp)$, the graphs contributing to $\b^{\a,1}_{\m,k}$ have at least
one propagator on scale $0$ or $-1$; by the short memory property,
this means that they can be dimensionally bounded by $(\const.)\bar\e_k^2
M^{\th k}$, for any $0<\th<1$. Similarly, the contributions to
$\b^{\a,2}_{\m,k}$ associated to graphs with at least one vertex of type
$\a_{\m,k'}$ for some $k'> k$ can be bounded by $(\const.)\bar\e_k^2|
\a_{\m,k'}|M^{\th (k-k')}$. The counterterms $\a_{\m}$ are fixed in such a way
that $\a_{\m,-\io} =0$, i.e., $\a_{\m} = -\sum_{k=-\io}^{0}(\b^{\a,1}_{\m,k}+
\b^{\a,2}_{\m,k})$. Finally, by using the fact that $|\b^{\a,1}_{\m,k}|\le
(\const.)\bar\e_k^2 M^{\th k}$ and $|\b^{\a,2}_{\m,k}|\le (\const.)\bar\e_k^2|
\a_{\m,k'}|M^{\th (k-k')}$, we find that $|\a_{\m,h}| \le (\const.)
\bar\e_{h^{*}}^{2}M^{(\th/2)h}$. This dimensional estimate on
$\a_{\m,h}$ easily implies the desired estimate on $R_\m(\kk,\pp)$
stated in (\ref{restN}) and we will not belabor the details here.

{\it Lowest order computation of $\a_\m$.}
At lowest order,
$\a_\m^{(2)}=-\sum_{k\le 0}\b^{\a,1,(2)}_{\m,k}$, where $\b^{\a,1,(2)}_{\m,k}$
is the one-loop contribution to $\b^{\a,1}_{\m,k}$. Moreover,
$\b^{\a,1,(2)}_{\m,k}=0$ for all $k\le -1$. Therefore, neglecting
higher order terms, we find $\a_\m^{(2)}=-\b^{\a,1,(2)}_{\m,0}$, that is
(see Fig.\ref{fig3.2}):
\begin{figure}[htbp]
\centering
\includegraphics[width=0.35\textwidth]{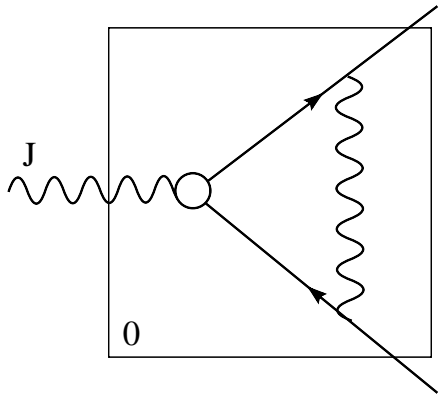}
\caption{Lowest order contribution to $\a_{0}$, $\a_{1}$.} \label{fig3.2}
\end{figure}
\bea &&\a_0^{(2)}=-i\g_0
\bar e_{\n,0}^{2}\int \frac{d\kk}{(2\pi)^3}\,
\g_{\n}\partial_{p_{0}}\Big[ g^{(0)}(\kk+\pp)C(\kk,\pp)g^{(0)}(\kk)\Big]_{\pp
=\V0}\cdot\nn\\&&\hskip1.5cm\cdot\g_{\n}w^{(0)}(\kk)\;,\label{A1.11}\\
&& \a_1^{(2)}= -\frac{i\g_1}{v}\bar e_{\n,0}^{2}\int \frac{d\kk}{(2\pi)^3}\,
\g_{\n}\partial_{p_{1}}\Big[ g^{(0)}(\kk+\pp)C(\kk,\pp)g^{(0)}(\kk)\Big]_{\pp
=\V0}\cdot\nn\\&&\hskip1.5cm\cdot\g_{\n}w^{(0)}(\kk)\,.\label{A1.11aa}\eea
After a straightforward computation, using the fact that
\bea
&&\dpr_{p_\m}\big[
g^{(0)}(\kk+\pp)C(\kk,\pp)g^{(0)}(\kk)\big]_{\pp=\V0}=\nn\\&&\hskip2cm=\frac1{i{\sl\kk}}
\big[-i\bar\g_\m\c_0(\kk)\big(1-\c_0(\kk)\big)+i{\sl \kk}\dpr_\m\c_0(\kk)\big]
\frac1{i{\sl\kk}}\;,\nn\eea
where ${\sl\kk}= k_0\g_0 + v\vec k\cdot \vec \g$ and $(\bar\g_0,\bar\g_1,
\bar\g_2)=(\g_0,v\g_1,v\g_2)$, we finally get (\ref{5.6b})-(\ref{5.7}).


\end{document}